\documentclass[aps,pra,groupedaddress,nofootinbib,notitlepage,showpacs,floatfix]{revtex4-1}
\usepackage{amsfonts}

\usepackage{graphicx,graphics,epsfig,times,bm,bbm,amssymb,amsmath,amsfonts,mathrsfs}
\usepackage[normalem]{ulem}
\usepackage{wrapfig}
\usepackage{boxedminipage}
\usepackage{setspace}
\usepackage{caption}
\usepackage{subfigure}
\usepackage{listings}
\usepackage{dsfont}
\usepackage{braket}
\usepackage[pdftex]{color}
\usepackage[pdfstartview=FitH]{hyperref}
\newcommand{\bes} {\begin{subequations}}
\newcommand{\ees} {\end{subequations}}
\newcommand{\bea} {\begin{eqnarray}}
\newcommand{\eea} {\end{eqnarray}}
\newcommand{\beq} {\begin{equation}}
\newcommand{\eeq} {\end{equation}}

\newcommand{\proj}[1]{\ketbra{#1}{#1}}

\newcommand{\om}{\omega}

\def\>{\rangle}
\def\<{\langle}
\def\Tr{\mathrm{Tr}}

\newcommand{\tr}{\text{tr}}

\newcommand{\Ei}{\text{Ei}}

\newcommand{\ketbra}[2]{|{#1}\>\<#2|}

\newcommand{\ident}{\mathds{1}}
\newcommand{\ignore}[1]{}

\newcommand{\reef}[1]{(\ref{#1})}

\begin{document}
\title{Coarse-Graining Can Beat the Rotating Wave Approximation in Quantum Markovian Master Equations}
\author{Christian Majenz$^{(1,3)}$}
\author{Tameem Albash$^{(2,3)}$}
\author{Heinz-Peter Breuer$^{(1)}$}
\author{Daniel A. Lidar$^{(2,3,4,5)}$}
\affiliation{$^{(1)}$Physikalisches Institut\\  Universit\"at Freiburg, Freiburg, Germany\\$^{(2)}$Department of Physics and Astronomy, $^{(3)}$Center for Quantum Information Science \&
Technology, $^{(4)}$Department of Electrical Engineering, $^{(5)}$Department of Chemistry\\University of Southern California, Los Angeles, California
90089, USA}
\begin{abstract}
We present a first-principles derivation of the Markovian semi-group master equation without invoking the rotating wave approximation (RWA). Instead we use a time coarse-graining approach which leaves us with a free timescale parameter, which we can optimize. Comparing this approach to the standard RWA-based Markovian master equation, we find that significantly better agreement is possible using the coarse-graining approach, for a three-level model coupled to a bath of oscillators, whose exact dynamics we can solve for at zero temperature.  The model has the important feature that the RWA has a non-trivial effect on the dynamics of the populations.  We show that the two different master equations can exhibit strong qualitative differences for the population of the energy eigenstates even for such a simple model. The RWA-based master equation misses an important feature which the coarse-graining based scheme does not. By optimizing the coarse-graining timescale the latter scheme can be made to approach the exact solution much more closely than the RWA-based master equation.
\end{abstract}
\maketitle

\section{Introduction}

The reduced dynamics of a quantum system coupled to another quantum system (the environment, or bath) can always be described by a completely positive trace preserving (CPTP) map provided the initial state of the system and the bath is factorized \cite{Breuer:2002}, or classically correlated \cite{1751-8121-41-20-205301,PhysRevLett.102.100402}.  The map can always be formally written in the Kraus operator sum representation (OSR):
\beq
\rho(t) = \sum_i A_i(t) \rho(0) A_i^{\dagger}(t) \ ,
\eeq
but calculating the Kraus operators $A_i$ explicitly is usually too difficult, making this form of the map unusable for most practical purposes.  It is often more convenient to work with a differential equation that describes the evolution.  Many such ``master equations'' have been derived, either from first principles or phenomenologically, and have been hugely successful in describing a wide range of phenomena \cite{Redfield,cohen1992atom,gardiner2004quantum,walls2008quantum,Breuer:2002}.  However, master equations in general do not necessarily generate CPTP dynamics, and can even violate the positivity of the density matrix for certain initial states.  Lindblad showed that a Markovian semi-group master equation (SME) that is guaranteed to generate a CPTP map must be of the form:
\beq \label{eqt:Lindblad}
\frac{d}{dt} \rho(t) = -\frac{i}{\hbar}\left[H, \rho(t) \right] + \frac{1}{2} \sum_{\alpha \beta} \gamma_{\alpha \beta}  \left(\left[L_{\alpha}, \rho(t) L_{\beta}^{\dagger} \right] + \left[L_{\alpha} \rho(t),  L_{\beta}^{\dagger} \right]  \right) \ , 
\eeq
where $H$ is Hermitian and the coefficients $\gamma_{\alpha \beta}$ form a positive matrix.  All first-principles derivations of the master equation that attempt to end in this form require what is often called the ``rotating wave approximation'' (RWA), also referred to as the ``secular approximation".  This approximation requires dropping terms which occur on timescales of order $1/(\omega - \omega') \ll \tau_R$, where $\omega$ and $\omega'$ are system energy eigenstate differences and $\tau_R \sim 1 / g$ is the typical relaxation time in the interaction picture, where $g$ is the system-bath interaction strength.  The RWA is notoriously hard to justify rigorously and is well known to sometimes give wrong results, e.g., Refs.~\cite{PhysRevLett.51.1108.3,PhysRevA.37.1628,PhysRevA.43.2430,Ford1997377,Fleming2010}. Nevertheless, it has been used extensively in the literature, and so it is tempting to formulate a general alternative approach to the first principles derivation of Markovian master equation.  

In Refs.~\cite{PhysRevA.60.1944,Lidar200135}, a (phenomenological) method was proposed to derive a Markovian SME from any CP map without requiring the RWA.  The procedure involves averaging the dynamics over a coarse-graining timescale. Therefore, we can derive two SMEs, one using the first-principles derivation with the RWA (which we refer to as the RWA-SME) and the second using the phenomenological coarse-graining idea of Refs.~\cite{PhysRevA.60.1944,Lidar200135} (which we refer to as the CG-SME).  In order to properly compare the effect of the RWA, we require a model to which an exact solution can be found, and in which the RWA acts non-trivially.  We construct a three-level system coupled to an infinite harmonic oscillator bath, the dynamics of which we can solve exactly at zero temperature.  Using this model, we compare the results of the two SMEs to the exact solution.  We find that even in this simple example, the dynamics of the RWA-SME misses an important qualitative feature of the populations of the exact solution, which the CG-SME does not. Moreover, no attempt has been made so far to optimize the coarse-graining timescale, and we show that the CG-SME can be fine-tuned to better match the exact solution via such an optimization.

The structure of this paper is as follows. In Section~\ref{sec:II} we first review the coarse-graining approach of Refs.~\cite{PhysRevA.60.1944,Lidar200135}, then derive a coarse-grained SME from first principles starting from the cumulant expansion approach of Alicki~\cite{PhysRevA.40.4077}. In Section~\ref{sec:trivial} we describe an exactly solvable 2-level model where the RWA-SME and CG-SME do not make substantially different predictions. Motivated by this observation, we construct and solve a 3-level model in Section~\ref{sec:model}, where the two master equations do make substantially different predictions, and derive the parameters of the two master equations in the context of the model in Section~\ref{sect:params}. We then compare the predictions to the exact solution in Section~\ref{sec:compare}, where we show that by optimizing the coarse-graining timescale we can get better agreement with the exact solution by using the CG-SME. In Section~\ref{sec:how} we briefly discuss the question of how to optimally choose the coarse-graining timescale in a more general setting, and conclude in Section~\ref{sec:conc}. Various technical details are presented in the Appendix.

%
\section{Semigroup master equations via coarse graining}
\label{sec:II}

\subsection{From CP maps to the SME} 
\label{sec:review}
%

In this subsection we briefly review the derivation of the SME \cite{PhysRevA.60.1944,Lidar200135}, to motivate the idea of time coarse-graining. We consider the dynamics of a system $S$ coupled to a bath $B$, together forming a closed system.  The closed system evolves unitarily under the Hamiltonian
\begin{equation}
H = H_S + H_B + H_{I} \ ,
\end{equation}
where $H_S$ is the system Hamiltonain, $H_B$ is the bath Hamiltonian, and $H_{I}$ is the interaction Hamiltonian.  Let us work in the interaction picture (we will denote interaction picture operators with a tilde (`` $\ \tilde{} \ $ ''), such that the total evolution operator is given by:
\beq
\tilde{U}(t) = \mathcal{T}_+ \exp \left[ - \frac{i}{\hbar} \int_0^{t} \tilde{H}_I(s) ds \right] \ ,
\eeq
where $\mathcal{T}_+$ denotes the time-ordering operator.  The system evolution is given by:
\begin{equation}
\tilde{\rho}_S(t) = \mathrm{Tr}_B \left[ \tilde{U}(t) \rho_S(0) \otimes \rho_B(0)  \tilde{U}^{\dagger}(t) \right] \ ,
\end{equation}
where we have assumed that at time $t = 0$ the closed system density matrix is factorized.  By writing the spectral decomposition in the basis $\{\ket{\mu}\}$ of the bath density matrix eigenstates,
\begin{equation}
\rho_B(0) = \sum_\mu \lambda_\mu | \mu \rangle \langle \mu | \ , \qquad  \lambda_\mu \geq 0 \ , \quad \sum_\mu \lambda_\mu = 1\ ,
\label{eq:rho_B}
\end{equation}
we can write the evolution equation in the operator-sum representation (OSR):
\begin{equation}
\tilde{\rho}_S(t) = \sum_{i} \tilde{A}_{i}(t) \rho_S(0) \tilde{A}_{i}^{\dagger}(t) \ ,
\end{equation} 
where the Kraus operators $\{\tilde{A}_i \}$ are given by 
\begin{equation} \label{eqt:Kraus}
\tilde{A}_i(t) = \sqrt{\lambda_{\nu}} \langle \mu | \tilde{U}(t) | \nu \rangle \ , \quad i = (\mu,\nu) \ .
\end{equation}
The Kraus operators have the property:
\begin{equation} \label{eqt:AA}
\sum_i \tilde{A}_i^{\dagger} \tilde{A}_i = \sum_{\mu,\nu} \lambda_\nu \langle \nu | \tilde{U}^{\dagger} | \mu \rangle \langle \mu | \tilde{U} | \nu \rangle  =  \sum_{\nu} \lambda_\nu = \ident \ .
\end{equation}

It is useful to expand the Kraus operators in terms of a fixed (time independent) operator basis $\{K_{\alpha} \}_{\alpha = 0}^{M}$ with $K_0 = \ident$ and $N_B$ the dimension of the bath Hilbert space, where $M = N_B^2-1$, such that:
\begin{equation}
\tilde{A}_i(t) = \sum_{\alpha = 0}^M b_{i \alpha}(t) K_{\alpha} \ ,
\end{equation}
to give:
\begin{equation} \label{eqt:1}
\tilde{\rho}_S(t) = \sum_{\alpha, \beta = 0}^{M} \chi_{\alpha \beta}(t) K_{\alpha} \rho_S(0) K_{\beta}^\dagger \ ,
\end{equation}
with $\chi_{\alpha \beta}(t) = \sum_{i = 1}^{N_B^2} b_{i \alpha}(t) b_{i \beta}^{\ast}(t)$.  In particular, $\chi_{\beta \alpha}^{\ast}(t) = \chi_{\alpha \beta}(t)$ and $\chi_{\alpha \alpha} = \sum_i |b_{i \alpha}|^2$, so $\chi$ is Hermitian with positive diagonal elements.  The property in Eq.~\eqref{eqt:AA} now gives
\begin{equation}
\sum_i \tilde{A}_i^{\dagger} \tilde{A}_i = \sum_{\alpha, \beta=0}^M \chi_{\alpha \beta} K_{\beta}^{\dagger} K_{\alpha} = \ident \ .
\end{equation}
By multiplying this expression from the left and right by $\rho(0)/2$ and adding the two expressions, we get the following relation:
\beq
\rho_S(0) = \chi_{00}(t) \rho_S(0) + \frac{1}{2} \sum_{\alpha=1}^M \chi_{\alpha 0}(t) \left\{K_{\alpha} , \rho_S(0) \right\}+\frac{1}{2}  \sum_{\beta= 1}^M \chi_{0 \beta}(t) \left\{\rho_S(0) , K_{\beta}^{\dagger}\right\} + \frac{1}{2} \sum_{\alpha, \beta = 1}^M \chi_{\alpha \beta}(t) \left\{ K_{\beta}^{\dagger} K_{\alpha}  , \rho_S(0) \right\} \ . \label{eqt:2}
\eeq
Taking the time derivative of Eq.~\eqref{eqt:1}, we are able to write an evolution equation:
\beq \label{eqt:Exact_Evolution}
\frac{d}{d t} \tilde{\rho}_S(t) =   - \frac{i}{\hbar} \left[ \dot{S}(t) , \rho_S(0) \right] + \frac{1}{2} \sum_{\alpha, \beta = 1}^{M} \dot{\chi}_{\alpha \beta}(t) \left( \left[ K_{\alpha}, \rho_S(0) K_{\beta}^{\dagger} \right] + \left[ K_{\alpha} \rho_S(0), K_{\beta}^{\dagger} \right] \right) \ ,
\eeq
where we have used the time derivative of Eq.~\eqref{eqt:2} to replace terms with $\dot{\chi}_{00}(t)$ and defined
\beq
S(t) = \frac{ i \hbar}{2} \sum_{\alpha = 1}^{M} \left( \chi_{\alpha 0}(t) K_{\alpha} - \chi_{0 \alpha}(t) K_{\alpha}^{\dagger} \right) \ .
\eeq

Although this form of the evolution equation has a striking resemblance to the SME in Eq.~\eqref{eqt:Lindblad}, it determines $\rho_S(t)$ in terms of $\rho_S(0)$, i.e. at a special time $t =0$ where the state is factorized into a product state.  To write an evolution equation such that the equation depends on $\rho_S(t)$, we must make some approximations.  Let us define three different time scales: (i) a time $\tau_c$ corresponding to the inverse of the high-frequency cut-off $\omega_c$ in the bath spectral density, (ii) a system time scale $\tau_S$ corresponding to the characteristic time for significant changes in the system density matrix in the interaction picture, (iii) and a coarse-graining time scale $\Delta t$ corresponding to the time-scale after which the bath has effectively ``reset.''  Let us assume the following relationship holds:
\beq
\tau_c \ll \Delta t \ll \tau_S \ .
\label{eq:timescales}
\eeq
Let us define $\rho_n = \tilde{\rho}(n \Delta t)$, such that discretizing Eq.~\eqref{eqt:Exact_Evolution} in terms of $\Delta t$ gives:
\beq
\frac{\rho_1 - \rho_0}{\Delta t} = - \frac{i}{\hbar} \left[ \langle \dot{S} \rangle , \rho_0 \right] + \frac{1}{2} \sum_{\alpha, \beta = 1}^{M} \langle \dot{\chi}_{\alpha \beta} \rangle \left( \left[ K_{\alpha}, \rho_0 K_{\beta}^{\dagger} \right] + \left[ K_{\alpha} \rho_0, K_{\beta}^{\dagger} \right] \right)  \ ,
\eeq
where
\beq
\langle X \rangle \equiv \frac{1}{\Delta t} \int_0^{\Delta t} X(s) ds \ .
\eeq
In principle, Eq.~\eqref{eqt:Exact_Evolution} cannot be used to determine the next discretization step relating $\rho_2$ in terms of $\rho_1$.  However, if we assume that the bath resets in the time $\Delta t$ such that the bath interacts with the system in \emph{exactly} the same manner at each time step $\Delta t$, then we can simply proceed with our discretization.  This amounts to performing a Markovian approximation, and we are led to the form of the SME:
\beq
\frac{d}{d t} \tilde{\rho}_S(t) =   - \frac{i}{\hbar} \left[ \langle \dot{S} \rangle , \tilde{\rho}_S(t) \right] + \frac{1}{2} \sum_{\alpha, \beta = 1}^{M} \langle \dot{\chi}_{\alpha \beta}\rangle \left( \left[ K_{\alpha}, \tilde{\rho}_S(t) K_{\beta}^{\dagger} \right] + \left[ K_{\alpha} \tilde{\rho}_S(t), K_{\beta}^{\dagger} \right] \right) \ .
\label{eq:CG-SME}
\eeq

It was shown in Ref.~\cite{Lidar200135} that the coefficients $\langle \dot{\chi}_{\alpha \beta}\rangle$ form a positive matrix, which means that Eq.~\eqref{eq:CG-SME} is in Lindblad form and hence generates completely positive dynamics. At no point was the RWA made in this derivation. 

Note further that the coarse-graining timescale $\Delta t$ enters Eq.~\eqref{eq:CG-SME} in a fundamental manner via the coarse-grained averages $\langle \dot{S} \rangle$ and $\langle \dot{\chi}_{\alpha \beta}\rangle$. Clearly, Eq.~\eqref{eq:timescales} leaves room for optimization of $\Delta t$, a point we shall return to in our discussion below.

\subsection{From the cumulant expansion to the SME}
%
We now present a semigroup master equation that can be derived from a first-principles theory, again without invoking the RWA, using coarse-graining instead. The derivation is somewhat lengthy, so we present it in four parts. We follow Alicki's work \cite{PhysRevA.40.4077} for the first two parts.  

\subsubsection{Cumulant expansion}
We first review the cumulant expansion approach of Ref.~\cite{PhysRevA.40.4077}. 
Let us consider a Hamiltonian:
\begin{equation}
H = H_S + H_B + \lambda H_I
\end{equation}
with
\begin{equation}
H_I = A \otimes B
\end{equation}
where $A$ is a Hermitian system operator and $B$ is a Hermitian bath operator.  We have restricted ourself to a single term to simplify the notation, but the more general case with multiple terms follows in an analogous fashion.  Let us assume that:
\begin{equation}
\langle B \rangle_B = 0 \  , \quad \rho(0) = \rho_S(0) \otimes \rho_B \ ,
\end{equation}
where 
\beq
\langle X \rangle_B \equiv {\rm Tr}[\rho_B X] ,
\eeq
and where $\rho_B$ is the Gibbs state for the bath:
\begin{equation}
\rho_B = \frac{e^{-\beta H_B}}{\mathcal{Z}_B} \ ,
\end{equation}
with $\mathcal{Z}_B = {\rm Tr}[\exp(-\beta H_B)]$ the partition function. Define:
\bes
\begin{align}
H_0(t) &\equiv H_S(t) \otimes I_B + I_S\otimes H_B \ , \\
U_0(t,0) &\equiv \mathcal{T} \exp \left( - i \int_0^t H_0(s) ds \right) \ , \\
\tilde{\rho}(t) &\equiv U_0^\dagger(t,0) \rho U_0(t,0) \ ,
\end{align}
\ees
where $\tilde{\rho}(t)$ is the state in the $H_0$-interaction picture.  In the interaction picture we have:
\begin{equation}
\frac{d}{dt} \tilde{\rho}(t) = - i \left[ U_0^{\dagger}  \lambda H_I  U_0 , \tilde{\rho} \right] \equiv - i \left[ \lambda \tilde{H}_I(t), \tilde{\rho} \right] \ ,
\end{equation}
with:
\begin{equation}
\tilde{H}_I(t) = U_{S}^\dagger(t) A U_{S}(t) \otimes U_B^\dagger(t) B U_{B}(t) \ , 
\end{equation}
where $U_{S}(t)$ is the unitary generated by $H_S$.
The formal solution is given by:
\begin{equation} \label{eqt:formal}
\tilde{\rho}(t) = \tilde{\rho}(0) - i \int_0^t ds \left[ \lambda \tilde{H}_I(s), \tilde{\rho}(s) \right] = \tilde{\rho}(0) - i \lambda \int_0^t ds \left[  \tilde{H}_I(s), \tilde{\rho}(0) \right] - \lambda^2 \int_0^t d s \int_0^s ds' \left[\tilde{H}_I(s), \left[ \tilde{H}_I(s') , \tilde{\rho}(0) \right] \right] + \dots \ .
\end{equation}
We are interested in the reduced density matrix:
\begin{equation}
\tilde{\rho}_S(t) = \Tr_B \left( \tilde{\rho}(t) \right) \equiv \tilde{W}_t \tilde{\rho}_S(0) \ .
\end{equation}
The cumulant expansion is given by
\begin{equation}
\tilde{W}_t = \exp \left( \sum_{n=1}^{\infty} \lambda^n K^{(n)}(t) \right) = I + \lambda K^{(1)}(t) + \lambda^2 \left( K^{(2)}(t) + \frac{1}{2} \left( K^{(1)} (t) \right)^2 \right) + O(\lambda^3)  \ ,
\end{equation}
and we solve for $K^{(n)}$ by matching powers of $\lambda$ with Eq.~\reef{eqt:formal}.  We get:
\begin{equation}
K^{(1)} (t) \tilde{\rho}_S(0) = - i \int_0^t ds \ \Tr_B \left( \left[ \tilde{H}_I^{(1)}(s) , \tilde{\rho}(0) \right] \right) = 0 \ ,
\end{equation}
where we have used $\langle B \rangle_B = 0$ (which can always be done by shifting the operator $B$ without loss of generality), and have used the shorthand:
\begin{equation}
A(t) = U_S^\dagger A U_S \ , \quad B(t) = U_B^\dagger B U_B  \ , \quad \tilde{H}_I^{(1)}(t) = A(t) \otimes B(t) \ .
\end{equation}
The next order in $\lambda$ gives:
\begin{equation}
K^{(2)} (t) \tilde{\rho}_S (0) =  - \int_0^t d s \int_0^s ds' \ \Tr_B \left( \left[ \tilde{H}_I^{(1)}(s), \left[ \tilde{H}_I^{(1)}(s') , \tilde{\rho}(0) \right] \right] \right) \ .
\end{equation}
(Note that our expression differs by a factor of $1/2$ in the second term relative to Eq.~(4.10) in Ref.~\cite{PhysRevA.40.4077}.)  We write:
\bes
\begin{eqnarray}
\label{eq:AArho}
 \Tr_B \left( \left[ \tilde{H}_I^{(1)}(s), \left[ \tilde{H}_I^{(1)}(s') , \tilde{\rho}(0) \right] \right] \right) &=& \left( A(s) A(s') \tilde{\rho}_S(0) - A(s') \tilde{\rho}_S(0) A(s) \right)  \Tr_B \left( B(s) B(s') \rho_B \right)  + \mathrm{h.c.} \\
 &=&  \left( A^\dagger(s) A(s') \tilde{\rho}_S(0) - A(s') \tilde{\rho}_S(0) A^\dagger(s) \right)  \Tr_B \left( B^\dagger(s) B(s') \rho_B \right)  + \mathrm{h.c.} 
\end{eqnarray}
\ees
A simple norm estimate shows that $\|K^{(n)}\| = O[(gt)^n]$, where $g = \|H_I\|$. Therefore the terms of third order and above can be neglected provided $\lambda g t \ll 1$. This amounts to making the Born approximation.

\subsubsection{The second order cumulant}

We restrict ourselves to the case of time-independent $H_S$, which was also considered in Ref.~\cite{PhysRevA.40.4077}. Here we give a much more detailed derivation.  Let
\begin{equation}
A(t) = U_S^\dagger(t) A U_S(t) = \sum_{a, b} e^{-i \left( \varepsilon_b - \varepsilon_a \right) t } |\varepsilon_a \rangle \langle \varepsilon_a | A | \varepsilon_b \rangle \langle \varepsilon_b| =  \sum_{\omega} A_{\omega} e^{-i \omega t } \ ,
\end{equation}
such that
\begin{equation}
A(s') = \sum_{\omega} A_{\omega} e^{- i \omega s' }  \ , \quad A^\dagger(s') = \sum_{\omega} A_{\omega}^\dagger e^{i \omega s' } \ . 
\end{equation}
This then gives the following map from time $0$ to $t$:
\beq
\label{eqt:TimeIndependent1}
K^{(2)} (t) \tilde{\rho}_S (0) = - \left[ \sum_{\omega, \omega'}\left(  A^\dagger_{\omega'} A_\omega \tilde{\rho}_S(0) - A_\omega \tilde{\rho}_S(0)  A^\dagger_{\omega'} \right)\mathcal{B}_{\omega \omega'}(t) + \mathrm{h.c.} \right] ,
\eeq
where
\bes
\bea
\label{eq:uneq}
\mathcal{B}_{\omega \omega'}(t) &\equiv& \int_0^t d s \int_0^s ds' e^{i ( \omega' s - \omega s' )}  \mathcal{B}(s,s') \ , \\
\mathcal{B}(s,s') &\equiv&  \langle B^\dagger(s) B(s') \rangle = \langle B(s) B(s') \rangle = \mathcal{B}(s',s)^\ast \ .
\eea
\ees

We show in Appendix~\ref{app:B-details} how Eq.~\eqref{eqt:TimeIndependent1} can be rewritten as
\begin{eqnarray}
\label{eq:K2-gen}
K^{(2)}(t) \tilde{\rho}_S(0) = - i \left[\mathcal{S}(t), \tilde{\rho}_S(0) \right] + \sum_{\omega, \omega'} b_{\omega \omega'}(t) \left[ A_{\omega} \tilde{\rho}_S(0) A_{\omega'}^\dagger - \frac{1}{2} \left\{A_{\omega'}^{\dagger} A_\omega , \tilde{\rho}_S(0) \right\} \right] \ ,
\end{eqnarray}
where
\beq\label{eq:bomompr}
b_{\omega \omega'} (t) \equiv \int_0^t d s \int_0^t ds'  e^{i ( \omega' s - \omega s' )}  \mathcal{B}(s,s') = b^{\ast}_{\omega' \omega}(t)\ .
\eeq
The ``Lamb shift" is
\begin{equation}
\mathcal{S}(t) =  \sum_{\omega, \omega'} S_{\omega \omega'}(t) A_{\omega'}^{\dagger} A_{\omega} ,
\end{equation}
where
\beq\label{eq:Lambshift}
S_{\omega \omega'}(t) = - \frac{i}{2} \int_0^t d s \int_0^s d s' \left( e^{i (\omega' s - \omega s') } \mathcal{B}(s,s') - e^{-i (\omega s - \omega' s') } \mathcal{B}(s',s) \right) .
 \eeq
Note that $\left(S_{\omega \omega'} \right)^{\ast} =  S_{\omega' \omega}$, so that $\mathcal{S}^\dagger = \mathcal{S}$, as required for the interpretation of $\mathcal{S}$ as a Hamiltonian.

\subsubsection{Complete positivity}

Clearly, the dissipative (second) term on the RHS of Eq.~\eqref{eq:K2-gen} appears to be in Lindblad form, but we must still prove the positivity of the matrix $b(t)$ (this was not done in Ref.~\cite{PhysRevA.40.4077}). To this end we again expand the bath density matrix in its eigenbasis [Eq.~\eqref{eq:rho_B}], and use this to write the correlation function $\mathcal{B}(s,s') = \< B^\dagger(s) B(s')\>_B$ explicitly. Let $\vec{v}$ be some arbitrary vector; then positivity amounts to showing that $\vec{v}b(t)\vec{v}^{\dagger} >0$ for all $\vec{v}$. Indeed:
\bes
\label{eq:bpos}
\begin{align}
\vec{v}b(t)\vec{v}^{\dagger} &= \sum_{\omega\omega'}v_\omega b_{\omega \omega'}(t) v^*_{\omega'} = \int_0^t d s \int_0^t ds'  \sum_{\omega}(v_\omega e^{-i \omega s' })  \sum_{\omega'}(v_{\omega'}e^{-i\omega' s})^* {\rm Tr}[\sum_\mu \lambda_\mu \ket{\mu}\bra{\mu} B^\dagger(s) B(s')] \\
& = \sum_\mu \lambda_\mu \bra{\mu}  F^\dagger(t) F(t)\ket{\mu} =  \sum_\mu \lambda_\mu \| F(t)\ket{\mu}\|^2 > 0 ,
\end{align}
\ees
where $F(t) \equiv \int_0^t ds B(s) \sum_\omega v_\omega e^{-i\omega s}$.

Note that we have arrived at Eq.~\eqref{eq:K2-gen} without making the Markov approximation.  Therefore, our (truncated) non-Markovian CP map is given by:
\beq \label{eqt:CPmap}
\tilde{\rho}_S(t) = e^{\lambda^2 K^{(2)}(t) } \rho_S(0)
\eeq
The complete positivity of this expression follows from the Lindblad theorem, as $K^{(2)}(t)$ is in Lindblad form for all $t > 0$.
\subsubsection{SME from the cumulant expansion and coarse-graining}
%
Finally, we show how to obtain the SME from the results above (this was not done in Ref.~\cite{PhysRevA.40.4077} either). Consider the CP map Eq.~\eqref{eqt:CPmap} and write it in the form of an evolution equation.  Expanding the exponential to second order in $\lambda$ and taking a time derivative, we have:
\beq
\frac{d}{dt} \tilde{\rho}_S(t) = \left[\lambda^2 \dot{\mathcal{S}}, \tilde{\rho}_S(0) \right] + \sum_{\omega, \omega'} \lambda^2 \dot{b}_{\omega, \omega'} \left[ A_{\omega} \tilde{\rho}_S(0) A_{\omega'}^{\dagger} - \frac{1}{2} \left\{ A_{\omega'}^{\dagger} A_{\omega}, \tilde{\rho}_S(0) \right\} \right] \ . 
\eeq
Written in this form, the equation looks identical in form to Eq.~\eqref{eqt:Exact_Evolution}, so we can make the identification:
\beq
\dot{S} \to \lambda^2 \dot{\mathcal{S}} \ , \quad \dot{\chi}_{\alpha \beta} \to \lambda^2 \dot{b}_{\omega \omega'} \ , \quad K_{\alpha} \to A_{\omega} \ , \quad K_{\beta}^{\dagger} \to A_{\omega'}^{\dagger}  \ .
\eeq
As discussed in section \ref{sec:review}, the path to a SME is simply to take the time average of $\dot{S}$ and $\dot{\chi}$ over the coarse-graining time $\Delta t$.  In particular, we have:
\bes
\begin{align}
\lambda^2 \langle \dot{\mathcal{S}} \rangle & =  \frac{\lambda^2}{\Delta t} \mathcal{S}(\Delta t) \equiv H'_{\mathrm{LS}}\ ,  \\
\lambda^2 \langle \dot{b}_{\omega,\omega'} \rangle & =  \frac{\lambda^2}{\Delta t} b_{\omega, \omega'}(\Delta t) \equiv \gamma_{\omega \omega'} \ .
\end{align}
\ees
The generator has the free parameter $\Delta t$ that has to be fixed by some means.  

It is useful at this point to compare the SME derived above with that typically derived using the RWA in the time-independent case.  For the latter case, which we refer to as the ``RWA-SME'', the SME takes the form \cite{Breuer:2002}:
\beq
\frac{d}{dt} \tilde{\rho}_S(t) = -i \left[ H_{\mathrm{LS}}, \tilde{\rho}_S(t) \right] + \sum_{\omega} \gamma({\omega}) \left( A_{\omega} \tilde{\rho}_S(t) A_{\omega}^\dagger - \frac{1}{2} \left\{A_{\omega}^{\dagger} A_{\omega}, \tilde{\rho}_S(t) \right\} \right) \label{eqt:SME_RWA}
\eeq
with
\bes
\label{eqt:RWA}
\begin{align} 
\gamma(\omega) &= \int_{-\infty}^{\infty} d s e^{i \omega s} \mathcal{B}(s,0) \ , \quad H_{\mathrm{LS}} = \sum_{\omega} S(\omega) A_{\omega}^{\dagger} A_{\omega}\ , \\   S(\omega) &= \int_{-\infty}^{\infty} d \omega' \gamma(\omega') \mathcal{P} \left(\frac{1}{\omega - \omega'} \right)\ .
\end{align}
\ees

In the former case, which we refer to as the ``CG-SME'', the system operators $A_\omega$ are the same
\beq \label{eqt:SME_Average}
\frac{d}{dt} \tilde{\rho}_S(t) = -i \left[ H'_{\mathrm{LS}}, \tilde{\rho}_S(t) \right] + \sum_{\omega, \omega'} \gamma_{\omega \omega'}(\Delta t) \left( A_{\omega} \tilde{\rho}_S(t) A_{\omega'}^\dagger - \frac{1}{2} \left\{A_{\omega'}^{\dagger} A_{\omega}, \tilde{\rho}_S(t) \right\} \right) \ ,
\eeq
 but the rates $\gamma$ keeps a dependence on two different eigenfrequencies $\omega$ and $\om'$:
\bes
\begin{align}
\label{eq:gammaww'}
\gamma_{\omega \omega'} (\Delta t) & =  \frac{\lambda^2}{\Delta t} b_{\omega \omega'} (\Delta t)   \ , \quad H'_{\mathrm{LS}} (\Delta t)=  \sum_{\omega,\omega'} S_{\omega \omega'} (\Delta t)A_{\omega'}^{\dagger} A_{\omega} \ , \\
 S_{\omega \omega'} (\Delta t)& =  - \frac{i}{2 \Delta t} \int_{0}^{\Delta t} ds \int_0^{s} d s'  \left( e^{i (\omega' s - \omega s')} \mathcal{B}(s,s') - e^{-i (\omega s - \omega' s')} \mathcal{B}(s',s) \right) \ .
 \end{align}
\ees

We show in Appendix \ref{appendix:RWA} that the RWA result is recovered by taking the $\Delta t \to \infty$ limit of the above result. This is not surprising, since the RWA (as it is performed in the standard first-principles derivation of the Lindblad SME) is equivalent to averaging the time-dependent ME operator over several periods of the oscillating terms. In Ref.~\cite{Lidar200135} it was argued via a spin-boson example that ultimately the CG-SME generator should not depend on the averaging time $\Delta t$, and this is true if $\Delta t$ is well beyond the bath correlation time $\tau_B$. This corresponds to the RWA limit as mentioned above (i.e., $\Delta t \to \infty$), at least for the SME generator that results from the CP-map derived from the cumulant expansion. As we desire to derive a master equation that accounts for dynamics slightly beyond the RWA regime, we do not take the RWA limit and keep the dependence on the coarse-graining time scale.

Our main focus will be a comparison of the predictions of Eqs.~\eqref{eqt:SME_RWA} (RWA-SME) and \eqref{eqt:SME_Average} (CG-SME), both to each other, and to the exact solution of a model we describe next.

\section{An Exactly Solvable Model with Trivial Rotating Wave Approximation} 
\label{sec:trivial}
%
Let us consider the problem of a single 2-level system coupled to a bath of harmonic oscillators:
\begin{equation}
H_S = \frac{\omega_0}{2} \sigma_z \ , \quad H_B = \sum_k \omega_k b_k^\dagger b_k  \ , \quad \lambda H_{SB} = \lambda \sum_k \sigma_z \left( b_k^\dagger + b_k \right) \ ,
\end{equation}
whose exact solution is known \cite{Breuer:2002}.  We wish to compare the exact solution with the solution found using the RWA-SME and the CG-SME.  In all three cases, the diagonal elements of the density matrix are independent of time, and the off-diagonal component takes the form:
\begin{equation}
\rho_{12}(t) = e^{- i \omega_0 t} e^{ \Gamma(t)} \rho_{12}(0) \ .
\end{equation}
For the three cases, the dephasing rate is given by:
\begin{eqnarray}
\Gamma_{\textrm{exact}}(t) &=& - 4 \lambda^2 \int_0^{\infty} d \omega J(\omega) \coth \left( \frac{\beta \omega}{2} \right) \left( \frac{ 1 - \cos(\omega t) }{\omega^2} \right) \\
\Gamma_{\textrm{CG}}(t,\Delta t) &=& - 4 \lambda^2 \frac{t}{\Delta t} \int_0^{\infty} d \omega J(\omega) \coth \left( \frac{\beta \omega}{2} \right) \left( \frac{ 1 - \cos(\omega \Delta t) }{\omega^2} \right) \\
\Gamma_{\textrm{RWA}}(t) &=& - 4 \pi \lambda^2 t \lim_{\omega \to 0} \left( \frac{ J(\omega)}{1- e^{-\beta \omega}} \right)
\end{eqnarray}
where $\Gamma_{\textrm{exact}}$ is the exact result, $\Gamma_{\textrm{CG}}$ is the CG-SME result, and $\Gamma_{\textrm{RWA}}$ is the RWA-SME result (details of the derivation are given in Appendix \ref{app:2level}).  Formally, $\Gamma_{\textrm{CG}}(t, t) = \Gamma_{\textrm{exact}}(t)$. However, recall that $\Delta t$ is supposed to be a {\em fixed} timescale, so setting $\Delta t=t$ isn't in the spirit of the CG approximation. Both the CG- and RWA-SME yield a linear time-dependence for the dephasing, which is to be expected because of the Markovian approximation.  Furthermore, the CG-SME reduces to the RWA-SME in the limit of $\Delta t \to \infty$ (also shown in Appendix \ref{app:2level}).  Since the SMEs have the exact same behavior for the populations of the two energy eigenstates as the exact solution, we refer to this as a trivial case of the RWA.

\section{Exactly Solvable Model with Nontrivial Rotating Wave Approximation}
\label{sec:model}
%
We have so far presented a derivation of a SME without the RWA.  In order to investigate the effect of taking/not-taking the RWA, we require an exactly solvable model that has more than one bath-induced transition. To this end, we consider a three level system (with energy eigenstates denoted by $| 0 \rangle$, $|1 \rangle$, and $|2 \rangle$, and with energies $0$, $\omega_1$ and $\omega_2$ respectively) 
\beq
H_S=\om_1 \ketbra{1}{1}_S+\om_2 \ketbra{2}{2}_S \label{eq:exham_2} \ .
\eeq
This system is also refered to as V-type three level system. This system is coupled to a bath of non-interacting harmonic oscillators at zero temperature:
\beq
H_B=\sum_{k}\om_k b_k^\dagger b_k \label{eq:exham_3} \ .
\eeq
We choose the system-bath interaction such that transitions between the second and the third level are forbidden:
\bes
\begin{align}
H_I =&\ketbra{1}{0}B_1+\ketbra{2}{0}B_2+\text{h.c.}\label{eq:exham_4} \ , \\
B_i =&\sum_k g_{ik}b_k \ . \label{eq:exham_5}
\end{align}
\ees
This is a generalization of the one-excitation example in Ref.~\cite{Breuer:2002}[p.461].
The ``number operator''
\beq
N=\ketbra{1}{1}+\ketbra{2}{2}+\sum_k b_k^\dagger b_k \ ,
\eeq
commutes with the closed system Hamiltonian $ H = H_S + H_B +H_I$  and therefore is a conserved quantity. 
Let us define $H_0=H_S+H_B$ and transform to the interaction picture:
\bes
\begin{align}
\tilde{H}_I(t)=&e^{i H_0 t}H_Ie^{-i H_0 t}\\
=&e^{i \om_1 t}\ketbra{1}{0}B_1(t)+e^{i \om_2 t}\ketbra{2}{0}B_2(t)+\text{h.c.} \ , \label{eq:intintham}
\end{align}
\ees
where
\beq
B_i(t)=\sum_k g_{ik}b_k e^{-i \om_k t} \ .
\eeq
As there are no excitations in the bath at zero temperature, we can restrict ourselves to the subspace with eigenvalues $N = 0 ,1$, which is spanned by $|0\>_{SB}=|0\>\otimes|0\>_B$, $|1\>_{SB}=|1\>\otimes|0\>_B$, $|2\>_{SB}=|2\>\otimes|0\>_B$ and $|k\>_{SB}=|0\>\otimes|k\>_B$.  Note that $| k \rangle_B$ corresponds to having a single excitation of the $k$-th oscillator.  Therefore, we can write the wavefunction at time $t$ as
\beq
|\psi(t)\>=c_0|0\>+c_1(t) |1\>+c_2(t)|2 \rangle+\sum_k c_k(t)|k \rangle \ ,
\eeq
where $c_0$ is time independent since $H_I(t)|0\>=0 \ ,  \forall t$. For later comparison with the master equation results we need the reduced density matrix
\beq
\tilde{\rho}_S=\tr_B(\proj{\psi})=\begin{pmatrix}1-|c_1|^2-|c_2|^2&c_0c_1^*&c_0c_2^*\\c_0^*c_1&|c_1|^2&c_1c_2^*\\c_0^*c_2&c_1^*c_2&|c_2|^2\end{pmatrix} \ ,
\eeq
where we omitted the time arguments for notational simplicity. The interaction picture Schr\"odinger equation
\beq
i\frac{d}{dt}|\psi(t)\>=\tilde{H}_I(t)|\psi(t)\> \ , 
\eeq
yields a set of coupled differential equations for the coefficients (time arguments omitted):
\bes
\begin{align}
\dot{c}_i=&-i\sum_kg_{ik}e^{i(\om_i-\om_k)t}c_k \ , \quad  i=1,2\label{eq:sysdegs} \ , \\
\dot{c}_k=&-i\left(g_{1k}^*e^{-i(\om_1-\om_k)t}c_1+g_{2k}^*e^{-i(\om_2-\om_k)t}c_2\right) \ . \label{eq:bathdegs}
\end{align}
\ees
We can integrate Eq.~\eqref{eq:bathdegs} and plug it into Eq.~\eqref{eq:sysdegs} which yields for $c_1$
\beq
\dot{c}_1=-\sum_k \left[\left|g_{1k}\right|^2\intop_0^td\tau e^{i(\om_1-\om_k)(t-\tau)}c_1(\tau)+g_{1k}g_{2k}^*e^{i(\om_1-\om_2)t}\intop_0^td\tau e^{i(\om_2-\om_k)(t-\tau)}c_2(\tau)\right] \ .
\eeq
Lets assume for simplicity that $g_{1k}=g_{2k}\equiv g_{k}$. Note that if $\om_1=\om_2$, the state $|-\>=\frac{1}{\sqrt{2}}\left(|1\>-|2\>\right)$ is a dark state, i.e. it is decoupled from the bath, as both transitions $0\to 1$ and $0\to 2$ are coupled to the same bath operator $B$ now, and the model reduces to the two level model from \cite{Breuer:2002}.  

For the bath of harmonic oscillators, we will assume a continuum limit with spectral function $J(\omega)$ and a coupling constant $g$:
\beq\label{eq:contlim}
\sum_k \left|g_{k}\right|^2\rightarrow g\intop_{-\infty}^\infty d\om J(\om) \  , 
\eeq
such that we can replace the sum over bath oscillators by an integral over the spectral function
\bes
\begin{align}
\dot{c}_1& =  -f_1*c_1-e^{i(\om_1-\om_2)t}f_2*c_2 \ , \label{eq:c1conv} \\
\dot{c}_2& =  -f_2*c_2-e^{i(\om_2-\om_1)t}f_1*c_1 \ , \label{eq:c2conv}
\end{align}
\ees
where $*$ denotes the convolution
\beq
(f*g)(t)=\intop_0^td\tau f(t-\tau)g(\tau) \ ,
\eeq
and
\beq
f_j(t)=e^{i\om_j t}\intop_{-\infty}^\infty d\om J(\om)e^{-i \om t}\equiv e^{i\om_j t}f(t) \ , 
\eeq
is the frequency shifted Fourier transform of the bath spectral function.  Eqs.~\eqref{eq:c1conv} and \eqref{eq:c2conv} can be solved numerically to give the exact solution for the time evolution of the system density matrix.   We use a fixed step size fourth order Runge-Kutta method and calculate the convolution integral using the Riemann sum approximation with the stepsize of the Runge-Kutta method as interval width. 

%
%
\begin{figure}[ht]
\subfigure[\ Populations, $|\om_1-\om_2|=10 g$]{\includegraphics[width=3.0in]{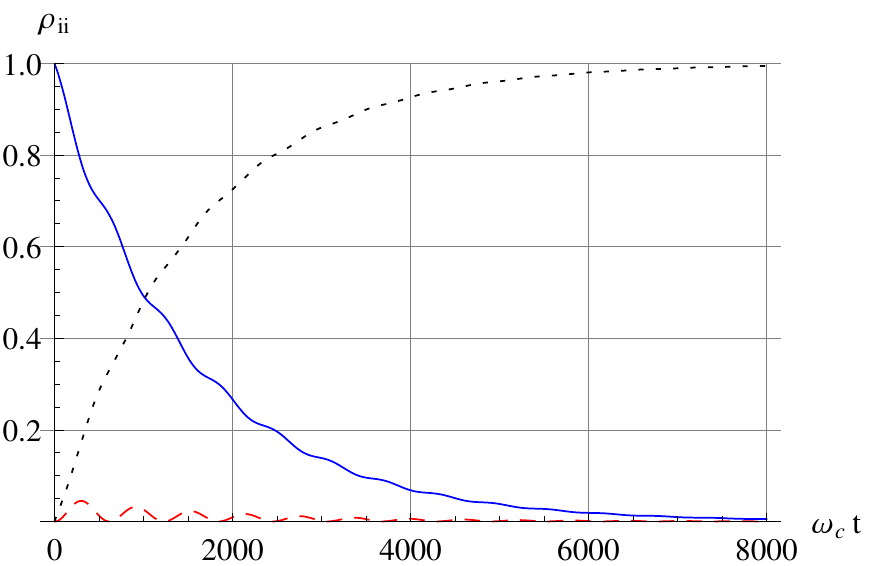} \label{fig:1a}} \hspace{0.5cm}
\subfigure[\ Populations, $|\om_1-\om_2|=\frac{g}{2}$]{\includegraphics[width=3.0in]{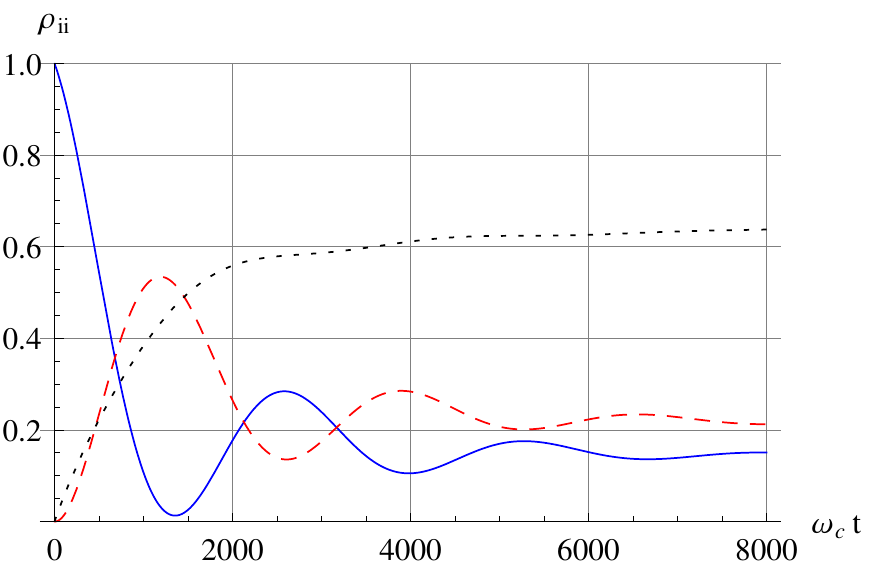} \label{fig:1b}} 
\subfigure[\ Coherence, $|\om_1-\om_2|=10 g$]{\includegraphics[width=3.0in]{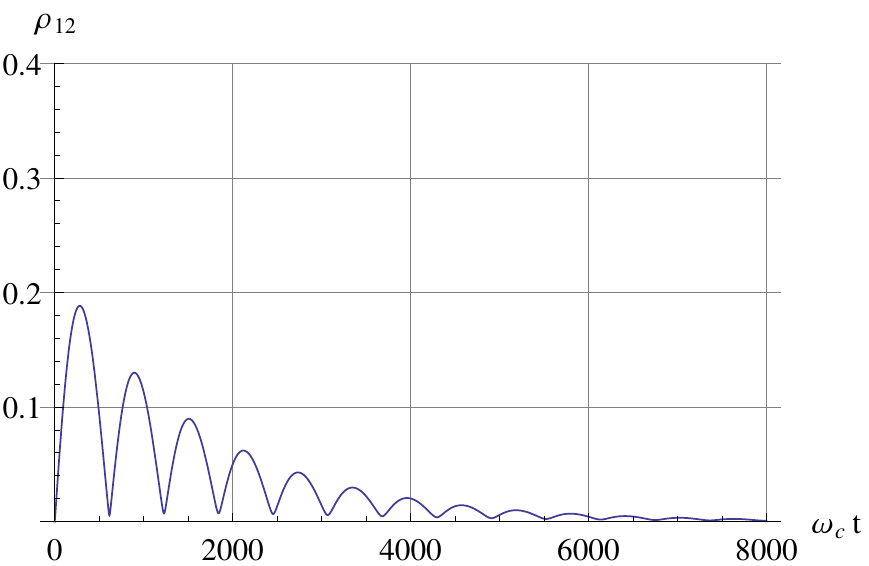}} \hspace{0.5cm}
\subfigure[\ Coherence, $|\om_1-\om_2|=\frac{g}{2}$]{\includegraphics[width=3.0in]{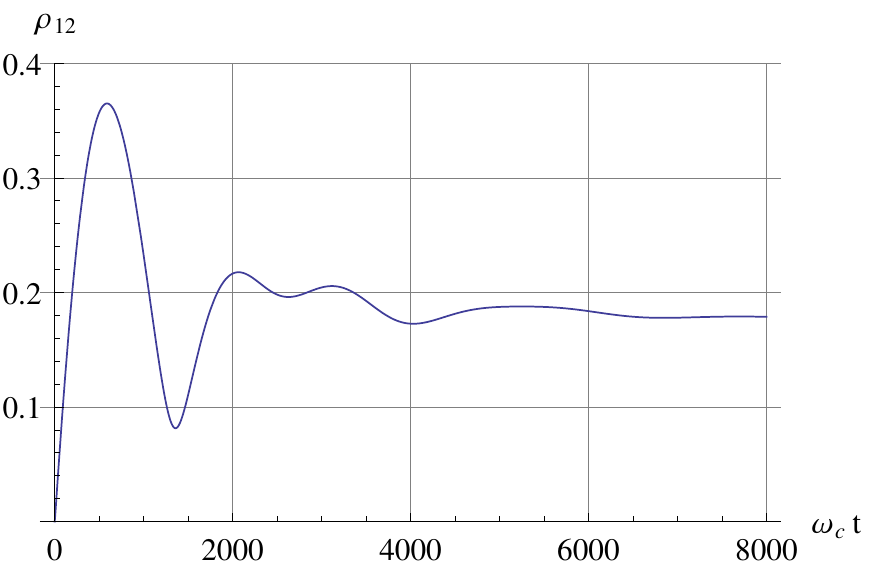}} 
\caption{Exact dynamics of the populations and coherences for $g=0.001\om_c$ and $\rho_S(0)=|1\>\<1|$.  In (a) and (c) we chose $\om_1=0.095\om_c$ and $\om_2=0.105\om_c$, in (b) and (d) $\om_1=0.09975\om_c$ and $\om_2=0.10025\om_c$. In (a) and (b) black and dotted is $\rho_{00}$, blue and solid is $\rho_{11}$ and red and dashed is $\rho_{22}$. In (c) and (d) we display the coherence between 
levels $1$ and $2$, i.e. $\rho_{12}$.}
   \label{fig:Exactdyn}
\end{figure}
\begin{figure}[ht]
\subfigure[\ CG-SME]{\includegraphics[width=3.0in]{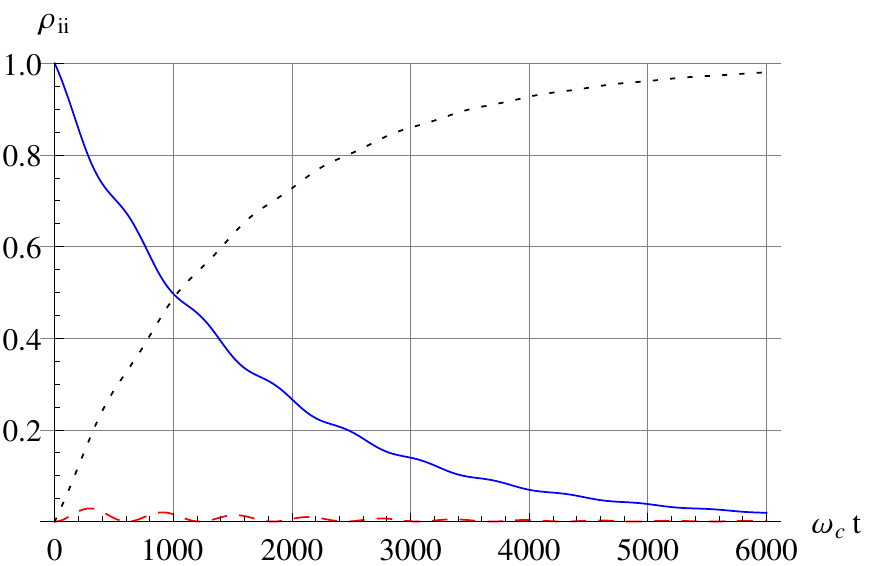} \label{fig:2a}} 
\subfigure[\ RWA-SME]{\includegraphics[width=3.0in]{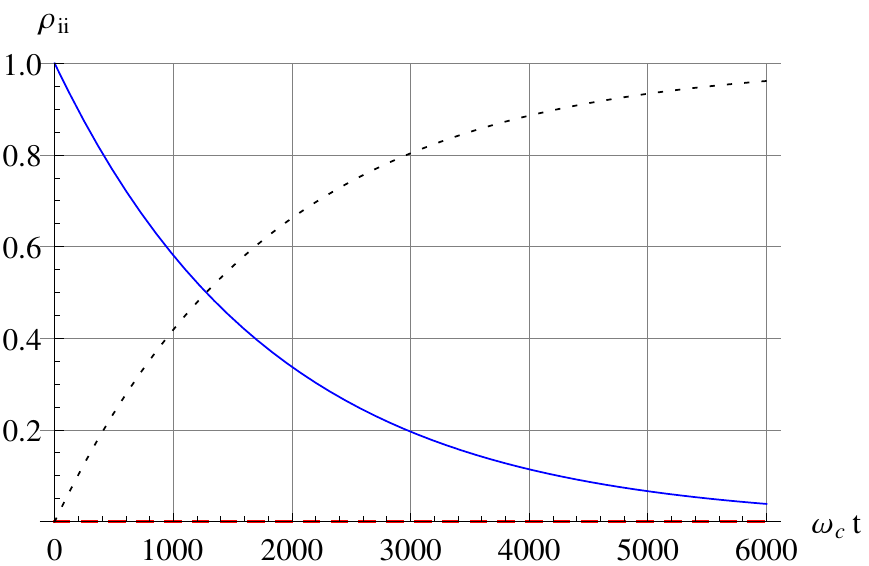}} \hspace{0.5cm}
\subfigure[\ Exact]{\includegraphics[width=3.0in]{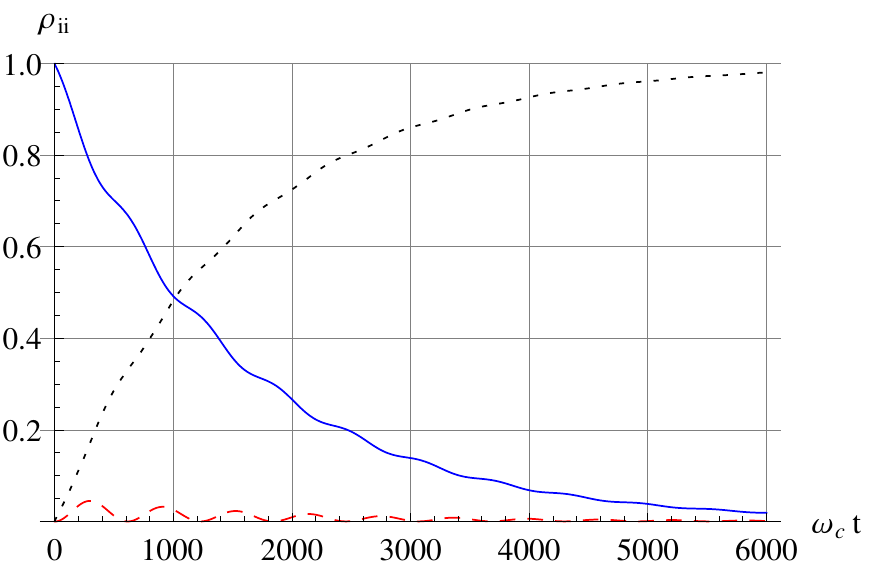} \label{fig:2c}} \hspace{0.5cm}
\caption{Dynamics of the populations $\rho_{ii}$ calculated with the different approximate equations compared to the exact dynamics at $g=0.001\om_c$, $\om_1=0.095\om_c$, $\om_2=0.105\om_c$. The dashed red line is $\rho_{22}(t)$, solid blue line is $\rho_{11}(t)$ and the black dotted line is $\rho_{00}(t)$. The optimal parameter for the averaged evolution generator is $\om_c\Delta t=63.7$ for this set of system parameters, in agreement with Eq.~\eqref{eq:timescales}. As initial condition we set $\rho_S(0)=|1\>\<1|$.}\label{fig:pops}
\end{figure}
\begin{figure}[h]
        \includegraphics*[angle=0,height=9cm]{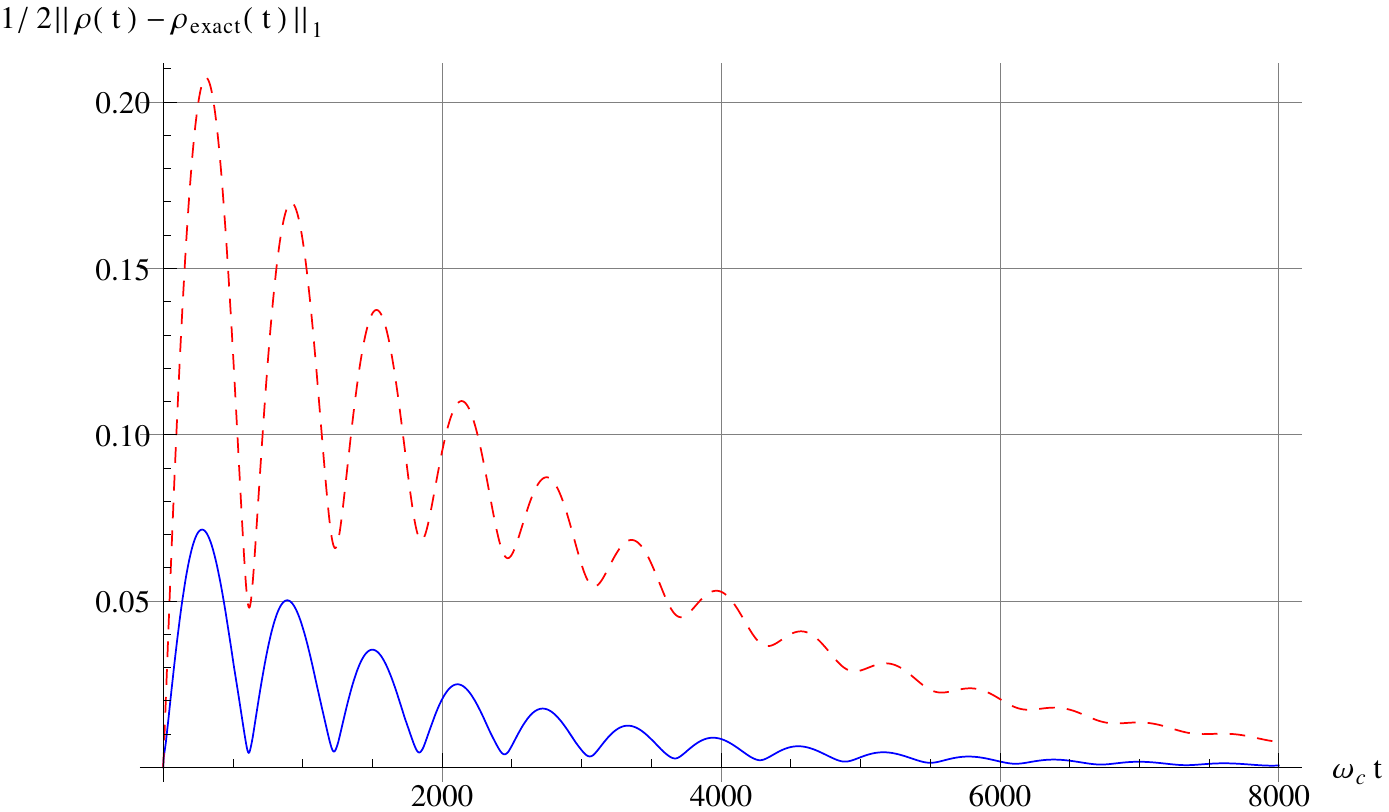}
        \caption{Trace-norm distance of the different approximate solutions to the exact solution, for the same paramters as in Fig. \ref{fig:pops}. The dashed red graph corresponds to the RWA-SME solution and the solid blue one to the CG-SME solution.}\label{fig:perf}
\end{figure}

\section{Parameters of the Master Equations}
\label{sect:params}
\subsection{RWA-SME}
We now wish to derive the SME parameters associated with the model described in section \ref{sec:model}.  First, we note that the interaction Hamiltonian in Eq.~\eqref{eq:exham_4} is of the form:
\beq
H_I = A \otimes B+  A^{\dagger} \otimes B^{\dagger} \ ,
\eeq
where
\beq
A = |1 \rangle \langle 0 | +  |2 \rangle \langle 0 | \ , \quad B =  \sum_k g_k  b_k \ .
\eeq
In the derivation of our SMEs it was assumed that $A$ and $B$ were Hermitian, but here they clearly are not.  This does not change the overall form of Eqs.~\eqref{eqt:SME_RWA} and \eqref{eqt:SME_Average}, but we must be careful how we define the bath correlation function.  Since we assume $T=0$, the bath state is $\rho_B=\ketbra{0}{0}$ and therefore we have the correlation functions (dropping the $B$ subscript for simplicity):
\bes
\begin{align}
\<B^\dagger(t)B(0)\>&=\<B^\dagger(t)B^\dagger(0)\>=\<B(t)B(0)\> = \<B^\dagger(0)B(t)\>=\<B^\dagger(0)B^\dagger(t)\>=\<B(0)B(t)\>=0 \ ,\\
\<B(t)B^\dagger(0)\>&=\sum_{k,k'}g_kg_{k'}^*\tr\left\{b_kb_{k'}^\dagger\ketbra{0}{0}\right\}e^{-i\om_k t} =\sum_{k}\left|g_k\right|^2e^{-i\om_k t}=\intop_{0}^{\infty} d \om J(\om)e^{-i \om t}=:\mathcal{B}(t,0)\label{eq:zeroTcorr} \ , \\
\<B(0)B^\dagger(t)\>&=\sum_{k,k'}g_kg_{k'}^*\tr\left\{b_kb_{k'}^\dagger\ketbra{0}{0}\right\}e^{i\om_k t} =\sum_{k}\left|g_k\right|^2e^{i\om_k t}=\intop_{0}^{\infty} d \om J(\om)e^{i \om t}=\mathcal{B}(t,0)^\ast \ ,
\end{align}
\ees
where we assumed the spectrum of the bath to be lower bounded by 0.  The one-sided Fourier transform of one of the nonzero correlation function is:
\bes
\begin{align}
\Gamma_+(\om)=&\intop_{0}^\infty d t\mathcal{B}(t,0)e^{i \om t}=\intop_{0}^\infty d \om'J(\om')\intop_0^\infty d t e^{-i t(\om'-\om)} =\intop_{0}^\infty d \om'J(\om')\left(\pi\delta(\om'-\om)+i \mathcal{P}\left(\frac{1}{\om-\om'}\right)\right) \\
=&\pi J(\om) \Theta(\omega)+i \intop_{0}^\infty d \om' J(\om') \mathcal{P} \left(\frac{1}{\om-\om'}\right) \label{eq:Gamma} \ ,
\end{align}
\ees
where we used the fact that
\beq
\intop_0^\infty d t e^{i \om t}=\pi \delta(\om)+i\mathcal{P}\left(\frac{1}{\om}\right) \ .
\eeq
$\mathcal{P}$ denotes the Cauchy principal value and and $\Theta$ denotes the Heaviside step function.  Similarly,
\beq
\Gamma_-(\omega)  = \intop_{0}^\infty d t\mathcal{B}(t,0)^\ast e^{i \om t} = \pi J(-\om) \Theta(-\omega)+i \intop_{0}^\infty d \om' J(\om') \mathcal{P} \left(\frac{1}{- \om-\om'}\right) = \Gamma_+(-\omega)^{\ast} \ .
\eeq
%
We assume our bath is characterized by an ohmic spectral density
\beq\label{eq:ohmic}
J_{\om_c}(\om)=\eta\om e^{-\frac{\om}{\om_c}} \ ,
\eeq
where $\eta$ has the dimension of time squared. We can calculate the integral in Eq.~\eqref{eq:Gamma} to give:
\bes
\begin{align}
\eta\intop_{0}^\infty d \om' \om'e^{-\frac{\om'}{\om_c}} \mathcal{P} \left(\frac{1}{\om-\om'} \right)=&-\eta\om_ce^{-\frac{\om}{\om_c}}\intop_{-\frac{\om}{\om_c}}^\infty d u \ e^{-u} \left( 1 + \frac{\omega}{\omega_c} \mathcal{P}\left(\frac{1}{u} \right) \right) \ , \\
=&-\eta\om_c+\eta\om e^{-\frac{\om}{\om_c}}\text{Ei}\left(\frac{\om}{\om_c}\right),
\end{align}
\ees
where we have used the exponential integral function:
\beq\label{eq:Ei}
\Ei (z)=-\intop_{-z}^\infty d\zeta e^{-\zeta} \mathcal{P} \left(\frac{1}{\zeta} \right) \ .
\eeq
The function has branch points at $z=0$ and $z=\infty$ and the branch cut is defined to go from $z=-\infty$ to $z=0$ .
Hence we get for the one-sided Fourier transform of the bath correlation function
\beq
\Gamma_+(\om)=\frac{1}{2}\gamma(\omega)+i S(\om)=\pi J_{\omega_c}(\om)+i\left[\eta\om_c+J_{\omega_c}(\om)\text{Ei}\left(\frac{\om}{\om_c}\right)\right].
\eeq
Similarly, 
\beq
\Gamma_-(- \omega) = \frac{1}{2} \gamma(\omega) - i S(\omega) \ .
\eeq
This allows us to write the generator from Eq.~\eqref{eqt:SME_RWA} as:
\beq
\mathcal{L} \tilde{\rho}_S(t) =-i\left[H_{\mathrm{LS}},\tilde{\rho}_S(t)\right]+\sum_{j=1,2}  \gamma(\om_j)\left(a_j \tilde{\rho}_S(t) a_j^\dagger-\frac{1}{2}\left\{a_j^\dagger a_j,\tilde{\rho}_S(t) \right\}\right) \ ,
\eeq
where
\beq
H_{\mathrm{LS}}=\sum_{j=1,2}  S(\om_j)a_j^\dagger a_j \ , 
\eeq
is the Lamb shift Hamiltonian and $a_i=\ketbra{0}{i}$. 
Transforming back to the Schr\"odinger picture just adds the system Hamiltonian to the unitary part.

%
\begin{figure}[h]
\subfigure[\, $\om=0.1\om_c$, $\delta \om=0.01\om_c$, $g=0.001\om_c$]{\includegraphics[width=3.0in]{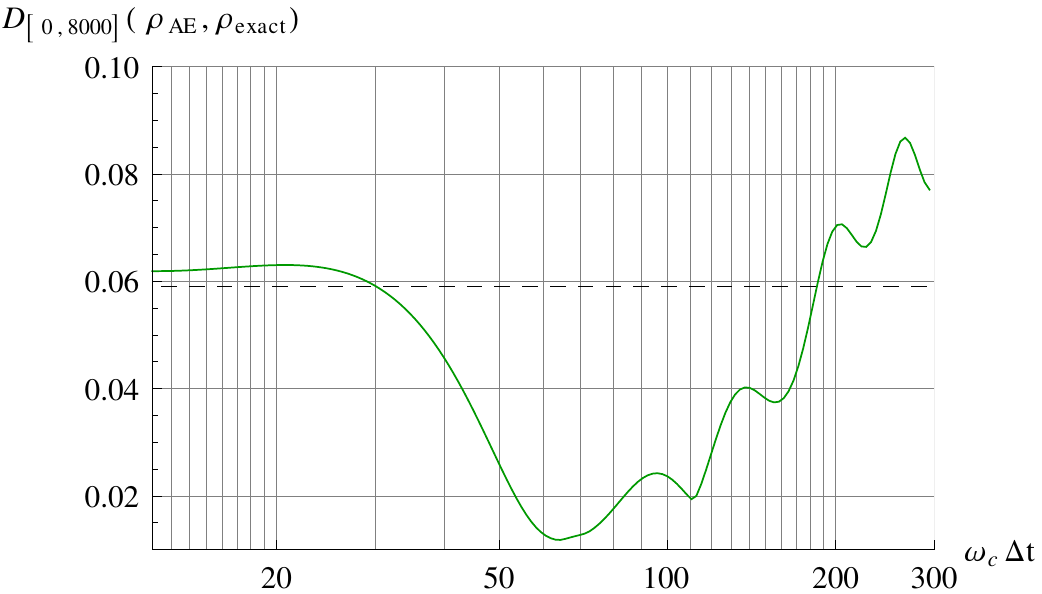}} 
\subfigure[\, $\om=0.1\om_c$, $\delta \om=0.01\om_c$, $g=0.002\om_c$]{\includegraphics[width=3.0in]{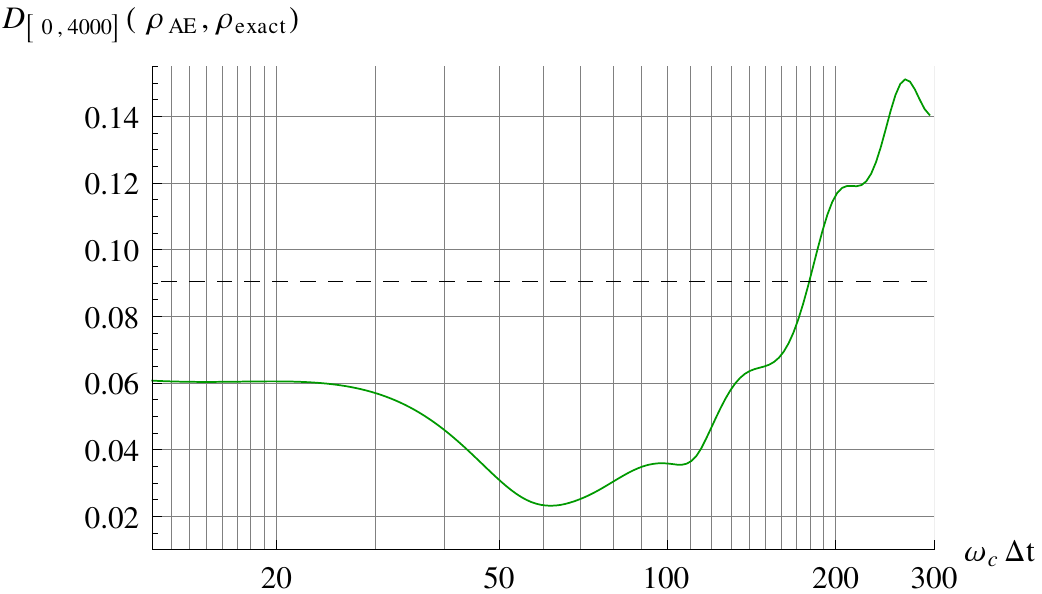}} \hspace{0.5cm}
\subfigure[\, $\om=0.1\om_c$, $\delta \om=0.01\om_c$, $g=0.003\om_c$]{\includegraphics[width=3.0in]{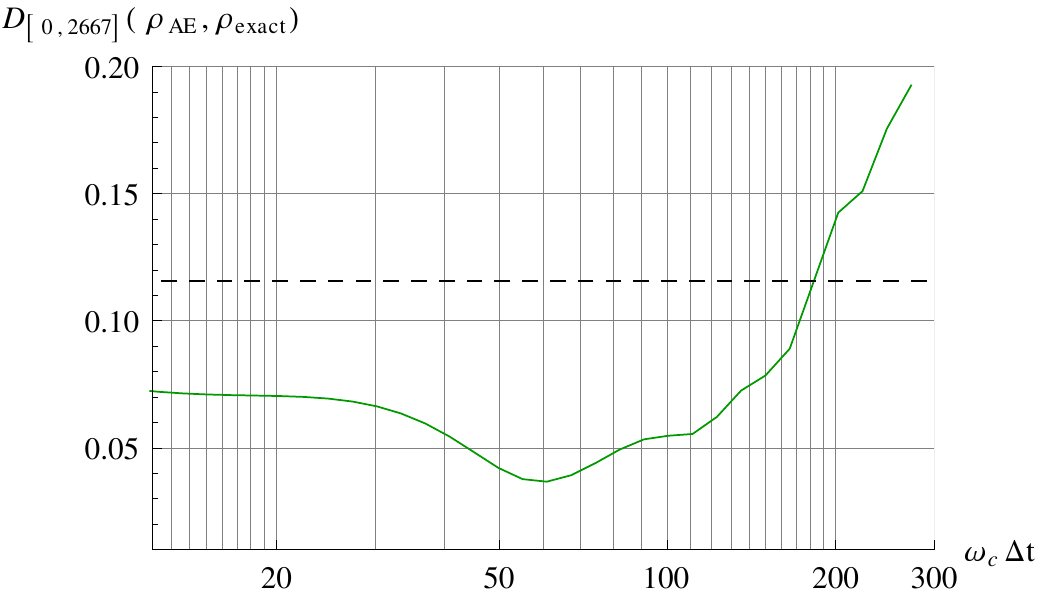}} 
\subfigure[\, $\om=0.4\om_c$, $\delta \om=0.01\om_c$, $g=0.001\om_c$]{\includegraphics[width=3.0in]{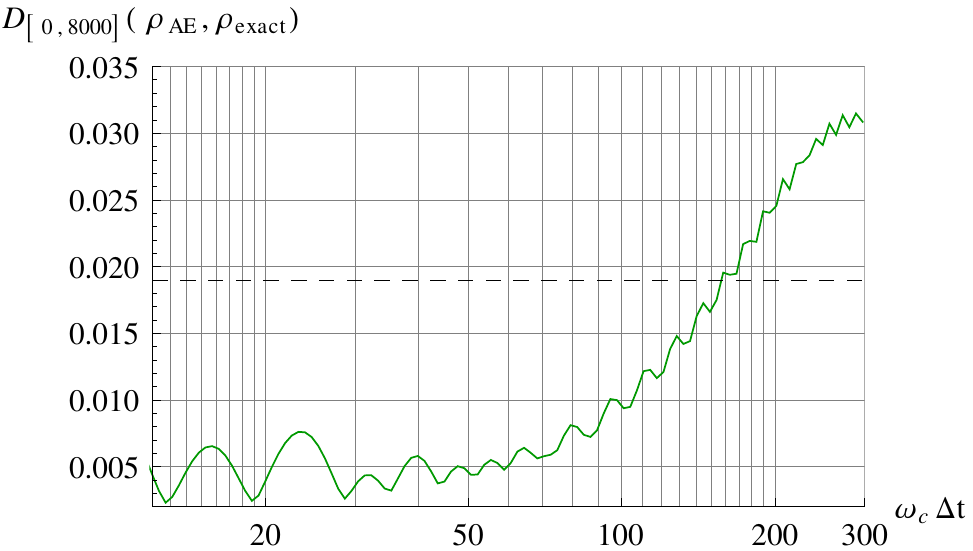}}\hspace{0.5cm}
\subfigure[\, $\om=0.3\om_c$, $\delta \om=0.03\om_c$, $g=0.003\om_c$]{\includegraphics[width=3.0in]{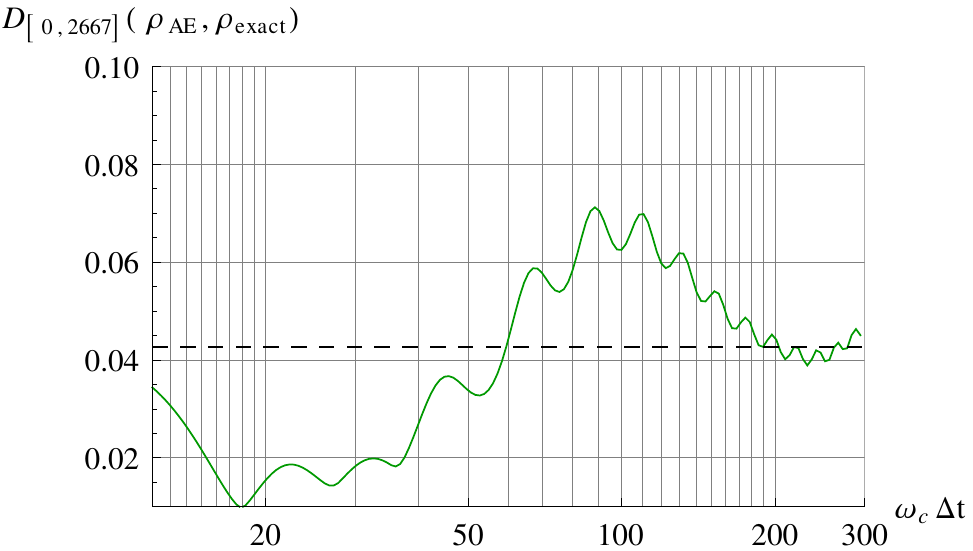}}
\caption{Integrated trace-norm distance of the solution of the CG-SME and the exact dynamics (green) for different averaging times $\Delta t$ compared to the RWA-SME (black, dashed). The integration intervals are chosen according to the relaxation times that result from the different system parameters.}\label{fig:ranges}
\end{figure}
\begin{figure}[h]
        \includegraphics*[angle=0,height=9cm]{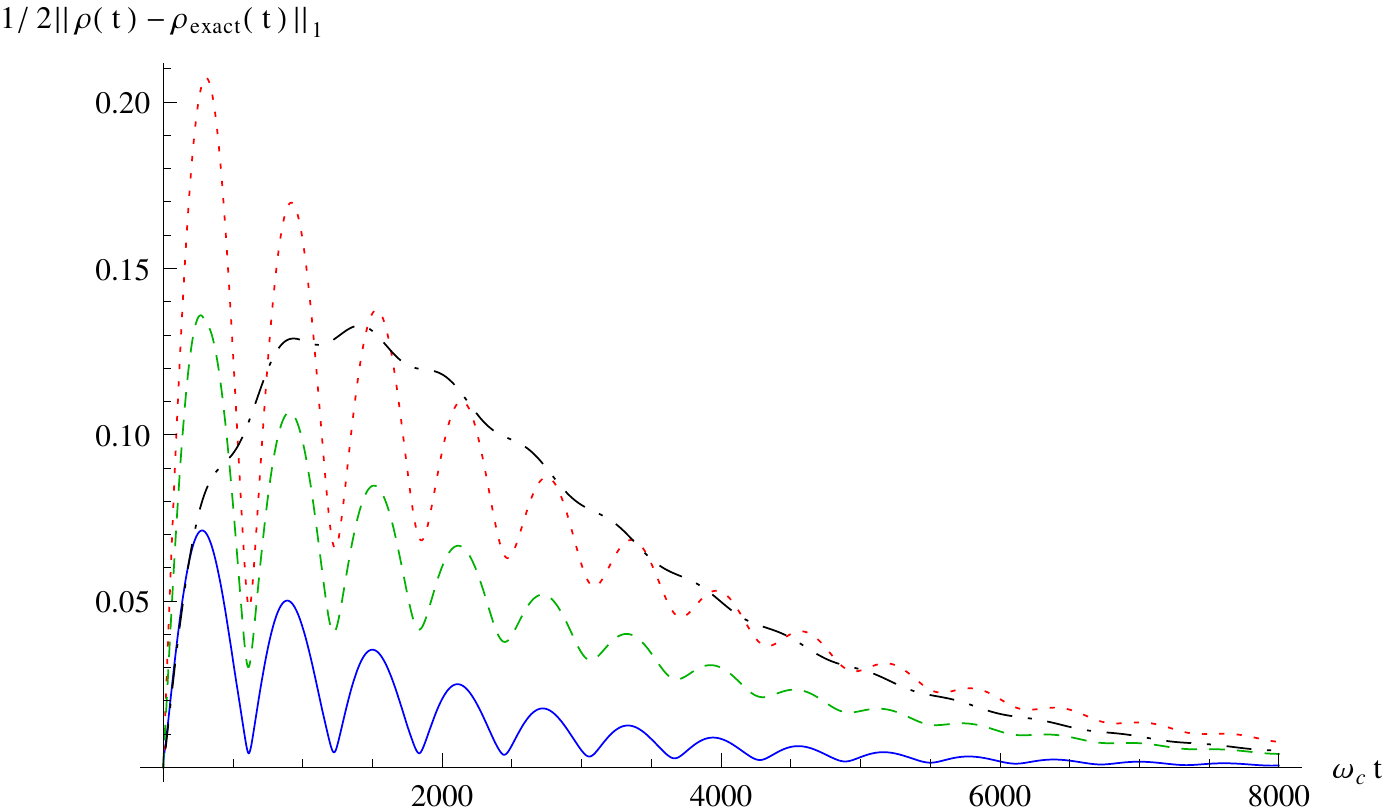}
        \caption{Trace-norm distance of the different approximate solutions to the exact solution, for the same paramters as in Fig. \ref{fig:pops}. The dotted red graph corresponds to the RWA-SME solution, the solid blue one to the optimal $\Delta t_{\textrm{opt}}$ CG-SME solution, and the dashed green and dashed-dotted black line are for $\Delta t=\Delta t_{\textrm{opt}}/2$ and $\Delta t=2\Delta t_{\textrm{opt}}$, respectively.}
        \label{fig:nonopttrdist}
\end{figure}
\begin{figure}[ht]
\subfigure[\ $\om=0.1\om_c$, $\delta \om=0.01\om_c$]{\includegraphics[width=3.0in]{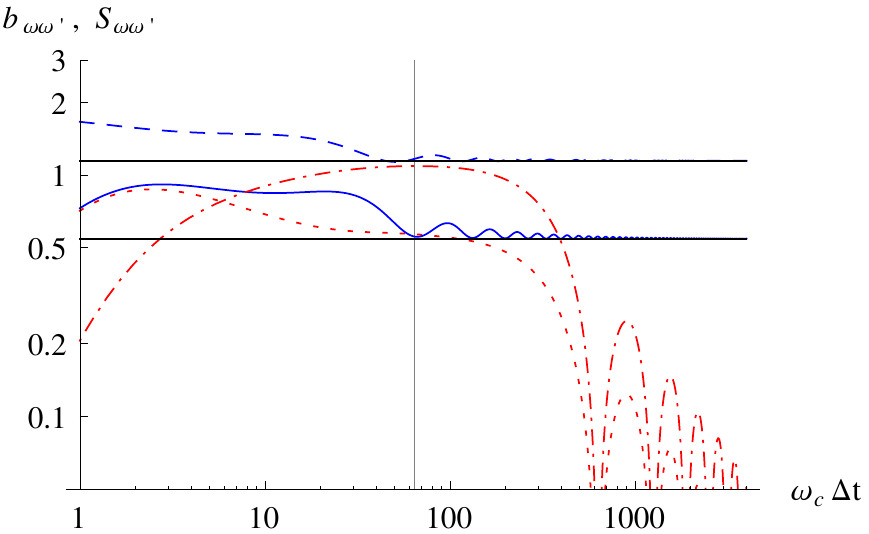}} 
\subfigure[\ $\om=0.4\om_c$, $\delta \om=0.01\om_c$]{\includegraphics[width=3.0in]{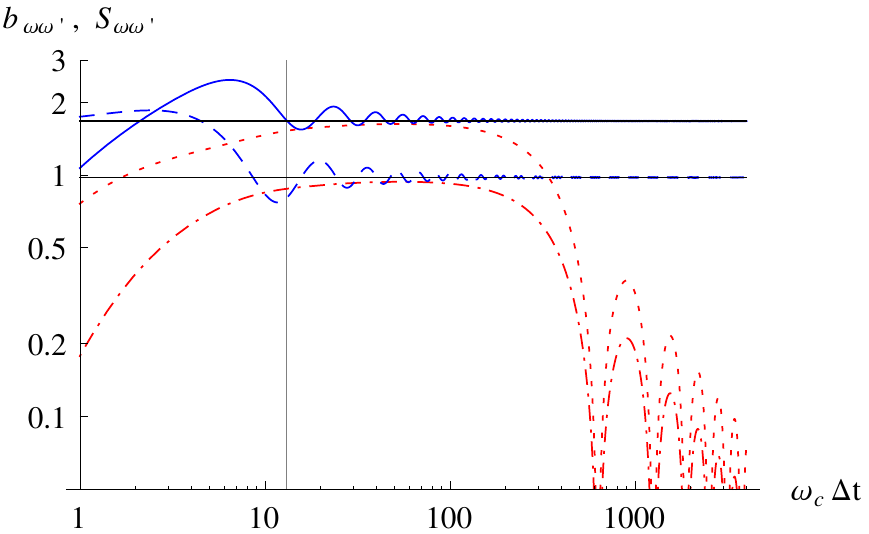}} \hspace{0.5cm}
\subfigure[\ $\om=0.2\om_c$, $\delta \om=0.02\om_c$]{\includegraphics[width=3.0in]{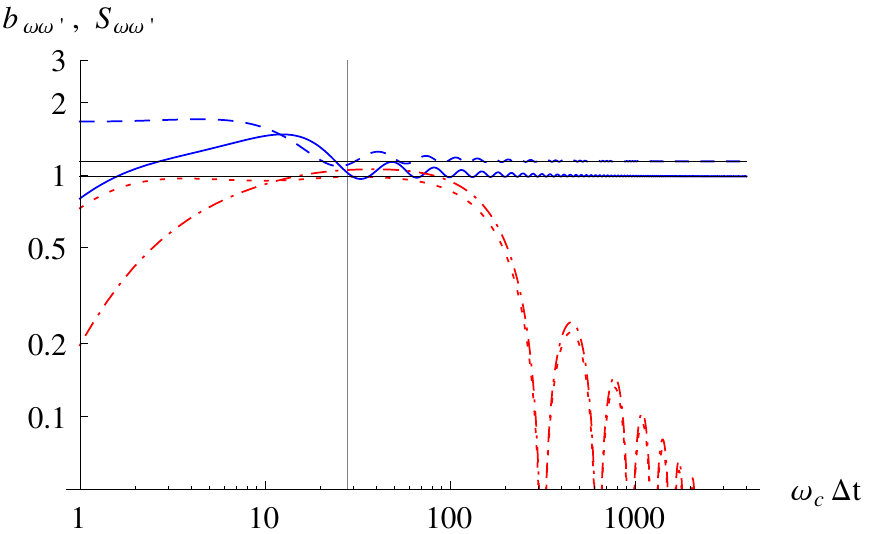}}
\caption{The rate functions of the CG-SME. The dotted red line is $b_{\om_1\om_2}$, the solid blue line $b_{\om_1\om_1}$, the dashed-dotted red line is $S_{\om_1\om_2}$ and the dashed blue line corresponds to $S_{\om_1\om_1}$. The horizontal lines are the asymptotic value of the rate functions, i.e., the value they converge to for high $\omega_c \Delta t$. The vertical lines are at $\om_c \Delta t_{\textrm{opt}}$. Note the qualitatively similar location of the $\om_c \Delta t_{\textrm{opt}}$ values relative to the features of the rate functions. This may be a clue for future studies in terms of independent determination of $\Delta t_{\textrm{opt}}$.}
\label{fig:rates}
\end{figure}

\subsection{CG-SME}
Here we construct the parameters of the CG-SME, Eq.~\eqref{eqt:SME_Average}. 
For the zero temperature case with the correlation function \eqref{eq:zeroTcorr}, Eq.~\eqref{eq:bomompr} simplifies to (for detailed calculations see Appendix \ref{sec:AEcalc})
\beq
b_{\omega \omega'}(t) = \intop_0^\infty d \nu \frac{J(\nu)}{(\om-\nu)(\om'-\nu)}\left(e^{i(\om'-\om)t}-e^{i(\om'-\nu)t}-e^{-i(\om-\nu)t}+1\right) \ .\label{eq:simpbwwpr}
\eeq
For $\om\not=\om'$ this yields
\beq 
b_{\omega \omega'}(t) = \frac{1}{\om'-\om}\left[\left(1+e^{i(\om'-\om)t}\right)\left(I_1(\om)-I_1(\om')\right)+e^{i\om' t}\left(I_2^*(\om')-I_2^*(\om)\right)+e^{-i\om t}\left(I_2(\om')-I_2(\om)\right)\right] \ ,
\eeq
where for the ohmic spectral density in Eq.~\eqref{eq:ohmic}: 
\bes
\label{eq:I1I2}
\begin{align}
\label{eq:I1}
I_1(\om)&=-g\om_c+g\om e^{-\frac{\om}{\om_c}}\text{Ei}\left(\frac{\om}{\om_c}\right) \\
\label{eq:I2}
I_2(\om)&=-\frac{g\om_c}{1-i\om_c t}+g\om e^{-\om\left(\frac{1}{\om_c}-i t\right)}\text{Ei}\left(\om\left(\frac{1}{\om_c}-i t\right)\right) \ .
\end{align}
\ees
For $\om=\om'$, Eq.~\eqref{eq:simpbwwpr} reduces to
\beq
b_{\om\om}(t)=2\intop_{-\om}^\infty d \nu \frac{J(\nu +\om)}{\nu^2}(1-\cos(\nu t)) \ .
\eeq
For the ohmic spectral density this yields
\bea
b_{\om\om}(t)&=&\eta e^{-\frac{\om}{\om_c}}\left[\left(1-\frac{\om}{\om_c}-i \om t\right)\Ei\left(\frac{\om}{\om_c}+i \om t\right)+\left(1-\frac{\om}{\om_c}+i \om t\right)\Ei\left(\frac{\om}{\om_c}-i \om t\right)+2\left(\frac{\om}{\om_c}-1\right)\Ei\left(\frac{\om}{\om_c}\right)\right] \nonumber \\
&& +2\eta(1-\cos(\om t)) .
\eea
Knowing the form of $b$ we can now compute the rates $\gamma$ in Eq.~\eqref{eq:gammaww'}. 

The Lamb shift term for the zero temperature bath correlation function in Eq.~\eqref{eq:zeroTcorr} is
\beq
S_{\om \om'}(t)=-\frac{ i}{2}\intop_0^\infty  d \nu J(\nu)\intop_0^t d s\left[\frac{e^{i s(\om'-\om)}-e^{i s(\om'-\nu)}}{i(\nu-\om)}+\frac{e^{i s(\om'-\om)}-e^{i s(\nu-\om)}}{i(\nu-\om')}\right] \ .\label{eq:Spartres}
\eeq
For $\om\not=\om'$ we get
\beq
S_{\om \om'}(t)=-\frac{i}{2(\om-\om')}\left[\left(1-e^{i t(\om'-\om)}\right)\left(I_1(\om)+I_1(\om')\right)+e^{-i\om t}\left(I_2(\om')-I_2(\om)\right)+e^{i\om' t}\left(I_2^*(\om)-I_2^*(\om')\right)\right] \ ,
\eeq
where we have again used the integrals defined in Eq.~\eqref{eq:I1I2}.  For $\om=\om'$, Eq.~\eqref{eq:Spartres} reads
\bea
S_{\om \om}(t)&=&- t I_1(\om)+I_3(\om) \ ,
\eea
where $I_1$ is defined in Eq.~\eqref{eq:I1} and $I_3$ is
\beq
I_3(\om)= \eta\frac{e^{-\frac{\om}{\om_c}}}{2 i}\left[\left(1-2\frac{\om}{\om_c}+2i\om t\right)\Ei\left(\frac{\om}{\om_c}+i\om t\right)-\left(1-2\frac{\om}{\om_c}-2 i\om t\right)\Ei\left(\frac{\om}{\om_c}-i\om t\right)\right]+\eta\sin(\om t) \ .
\eeq
We now have everything we need to construct the CG-SME, Eq.~\eqref{eqt:SME_Average}.

\section{Comparison between the master equations and the exact dynamics}
\label{sec:compare}

We first discuss the exact dynamics of the model presented in section \ref{sec:model}. For the calculations we set $\eta=\om_c^{-2}$. Fig.~\ref{fig:Exactdyn} shows two examples of the dynamics for different system parameters.  Since $[H_S,H_I]\neq 0$ for $|\om_1 - \om_2| \neq 0$,\footnote{Recall that as already pointed out in Section~\ref{sec:model}, for $\om_1=\om_2$ we have a dark state that is decoupled from the bath.} the dynamics are dissipative, so the system will eventually relax into the $T=0$ Gibbs state $\rho_{\text{Gibbs}}=|0\>\<0|$.  For  $|\om_1-\om_2|\gg g$, the system relaxes into the $T=0$ Gibbs state very rapidly [see Fig.~\ref{fig:Exactdyn}(a)].  For $|\om_1-\om_2|\lesssim g$, two timescales can be observed: the initial population of $|+\>=\frac{1}{\sqrt{2}}\left(|1\>+|2\>\right)$ decays rapidly, after that a strong oscillatory (with slow decay) behavior is observed in the $|1 \rangle$ and $|2 \rangle$ populations [see Fig.~\ref{fig:Exactdyn}(b)].  

We are interested in the regime where the RWA is no longer valid.  Recall that the RWA is argued to hold when the time scale associated with $|\omega_1 - \omega_2|^{-1}$ is much less than the relaxation time scale $\tau_R \sim 1 / g$.  Therefore, the RWA should start to break down when $|\om_1-\om_2|\gtrsim g$.  We investigate the behavior of our two SMEs in this regime.  
We choose the coarse-graining time scale $\Delta t$ such that it minimizes the integrated trace-norm distance between the CG-SME density matrix $\rho_{\textrm{CG}}$ and the exact dynamics $\rho_{\textrm{exact}}$:  
\beq
\mathcal{D}_{[0,t_{\max}]}(\rho_{\textrm{CG}},\rho_{\textrm{exact}})=\frac{1}{2 t_{\max}}  \intop_0^{t_{\max}} d t  \|\rho_{\textrm{CG}}(t)-\rho_{\textrm{exact}}(t)\|_1 \ , 
\eeq
where $\|\cdot \|_1$ is the Schatten 1-norm ($\|A\|_1 := {\rm Tr}\sqrt{A^\dag A}$).  In Fig.~\ref{fig:pops}, we show an example of the dynamics of the different master equations compared to the exact dynamics.  In this example, we consider an initial state given by $ \rho_S(0) = |1\rangle \langle 1|$.  The RWA-SME [Eq.~\eqref{eqt:SME_RWA}] shows no population in the state $|2 \rangle$ since the Lindblad operators only allow for transitions between $| 1 \rangle \rightarrow |0 \rangle$ and $| 2 \rangle \rightarrow |0 \rangle$.  This is in stark contrast to the CG-SME [Eq.~\eqref{eqt:SME_Average}] which properly shows the oscillatory population in the state $|2 \rangle$.  We can compare the results quantitatively by computing the trace distances to the exact result, shown in Fig.~\ref{fig:perf}.  The two SMEs show similar qualitative time behavior in the trace distance, but the CG-SME outperforms the RWA-SME.  
%
%
%
We calculate the optimal $\Delta t$ for the three level system for different sets of parameters (shown in Table \ref{opttable}):\\
\begin{table}
\caption{Optimal $\Delta t$ for different parameters.}	
\begin{center}
\begin{tabular}{c|c|c||c}\label{opttable}
	$\frac{\om}{\om_c}:=\frac{\om_1+\om_2}{2\om_c}$&$\frac{\delta\om}{\om_c}:=\frac{\om_2-\om_1}{\om_c}$&$g/\om_c$&$\om_c\Delta t_{\text{\textrm{opt}}}$\\\hline
	0.05&0.01&0.001&124\\
	0.1&0.01&0.001&63\\
	0.1&0.01&0.002&61\\
	0.1&0.01&0.003&59\\
	0.15&0.01&0.001&39\\
	0.2&0.02&0.002&28\\
	0.3&0.03&0.003&18\\
	0.4&0.04&0.004&13\\
	0.4&0.01&0.001&13
\end{tabular}
\end{center}
\end{table}
%

\section{How can we find a good $\Delta t$ for real problems?}
\label{sec:how}
If we want to make the CG-SME useful more generally, we need to find a way of determining $\Delta t$ without comparing it to an exactly calculated result. 
We calculate the integrated trace-norm distance of the solutions of the CG-SME and the exact solution for suboptimal values adjacent to the optimum, shown in Fig.~\ref{fig:ranges}.

The numerical result suggests that for this particular three-state model, the optimal averaging time $\Delta t$ strongly depends on the timescale $\om^{-1}$ of the transitions driven by the bath, and does not depend strongly on the other parameters $g$ and $\delta \omega$. The latter are the ones that determine the validity of the RWA. On the other hand we can see, that if the RWA fails, i.e. if the RWA-SME solution has significant trace-norm distance from the exact solution, there is a wide range of values for the averaging time where the CG-SME does not yield optimum results but still outperforms the RWA-SME.  This fact is further clarified in Fig. \ref{fig:nonopttrdist}, where using half the optimal value as well as its double  still outperform the RWA-SME on average.

It is instructive to look at the rates of the CG-SME for different $\Delta t$ (Fig. \ref{fig:rates}). The optimal value of $\Delta t$ is located at a similar point on (the similar looking) graphs, independent of the parameters of the model. This fact might make an analytical determination of the optimal averaging time possible, but we leave the investigation of such a possibility to future research.
%
\section{Conclusions}
\label{sec:conc}
%
First-principles derivations of quantum Markovian master equations have so far relied on the rotating wave approximation. Here we have shown how the RWA can be avoided in a first-principle derivation, and replaced by a coarse-graining procedure. The coarse graining timescale is a free, phenomenological parameter which is \textit{a priori} bounded between the bath correlation time (inverse of the high-frequency cut-off in the bath spectral density) and the characteristic timescale for significant changes in the system density matrix in the interaction picture, and can be optimized.  To test the new, coarse-graining based SME, we applied it to a simple three-state model coupled to an infinite harmonic oscillator bath, whose exact dynamics we can solve for.  Comparing, we found that the standard RWA-SME misses important qualitative features of the solution, while the CG-SME captures these features relatively well. By optimizing the coarse-graining timescale we achieved significantly better agreement with the exact solution by using the CG-SME rather than the RWA-SME. To perform the optimization we required the exact solution; however, we showed that there is a substantial range of the coarse-graining timescale parameter that still allows the CG-SME to outperform the RWA-SME. Future work should address how to extract reasonable and advantageous estimates of the coarse-graining timescale without relying on exact solutions.

\acknowledgments
C.M. is grateful to the USC Center for Quantum Information Science \& Technology for its hospitality. D.A.L. was supported by the ARO MURI grant
W911NF-11-1-0268, by NSF under grant numbers CHM-1037992, PHY-
969969, and PHY-803304.


\appendix

\section{From Eq.~\eqref{eqt:TimeIndependent1} to Eq.~\eqref{eq:K2-gen}}
\label{app:B-details}

The unequal upper integration limits in Eq.~\eqref{eq:uneq} are inconvenient. To remove them we note the following relations for the integral, where for notational simplicity we suppress the $t$-dependence for now:
\bes
\begin{align}
\mathcal{B}_{\omega \omega'} \equiv \int_0^t d s \int_0^s ds' e^{i ( \omega' s - \omega s' )}  \mathcal{B}(s,s') \ , =&\left[ \int_0^t d s \int_0^t ds' - \int_0^t d s \int_s^t ds'  \right] e^{i ( \omega' s - \omega s' )}  \mathcal{B}(s,s') \ ,  \\
=&  \left[\int_0^t d s \int_0^t ds' -  \int_0^t d s' \int_0^{s'} ds  \right] e^{i ( \omega' s - \omega s' )}  \mathcal{B}(s,s') \ , \\
=& \int_0^t d s \int_0^t ds'  e^{i ( \omega' s - \omega s' )}  \mathcal{B}(s,s')  -  \int_0^t d s \int_0^{s} ds'  e^{i ( \omega' s' - \omega s)}  \mathcal{B}(s',s) \ , \\
=& \ b_{\omega \omega'} - \left( \mathcal{B}_{\omega' \omega} \right)^{\ast} \ ,
\end{align}
\ees
where $b_{\omega \omega'}$ [Eq.~\eqref{eq:bomompr}] has the desired equal upper integration limits. It follows immediately that
\beq
\left( \mathcal{B}_{\omega \omega'}  \right)^{\ast} = b_{\omega' \omega} - \mathcal{B}_{\omega' \omega} \ .
\eeq
and from Eq.~\eqref{eq:bomompr} we also directly get
\begin{eqnarray}
\mathcal{B}_{\omega \omega'}&=& \frac{1}{2} b_{\omega \omega'} + \frac{1}{2} \left(  \mathcal{B}_{\omega \omega'}- \left( \mathcal{B}_{\omega' \omega} \right)^{\ast} \right) \ .
\end{eqnarray}
Therefore, Eq.~\eqref{eqt:TimeIndependent1} yields terms such as:
\bes
\begin{align}
\sum_{\omega, \omega'} \left[A_\omega \tilde{\rho} A_{\omega'}^\dagger \mathcal{B}_{\omega \omega'} + A_{\omega'} \tilde{\rho} A_{\omega}^\dagger \left(\mathcal{B}_{\omega \omega'} \right)^{\ast} \right] =& \sum_{\omega, \omega'} \left[A_{\omega} \tilde{\rho} A_{\omega'}^\dagger b_{\omega \omega'} + A_{\omega'} \tilde{\rho} A_{\omega}^\dagger b_{\omega' \omega} - \left( A_{\omega} \tilde{\rho} A_{\omega'}^\dagger \left(B_{\omega' \omega} \right)^{\ast} + A_{\omega'} \tilde{\rho} A_{\omega}^\dagger B_{\omega' \omega}\right) \right]  \ , \\
=&  \sum_{\omega, \omega'} \left[A_{\omega} \tilde{\rho} A_{\omega'}^\dagger b_{\omega \omega'} + A_{\omega'} \tilde{\rho} A_{\omega}^\dagger b_{\omega' \omega} - \left( A_{\omega'} \tilde{\rho} A_{\omega}^\dagger \left(B_{\omega \omega'} \right)^{\ast} + A_{\omega} \tilde{\rho} A_{\omega'}^\dagger B_{\omega \omega'}\right) \right] \ ,
\end{align}
\ees
where in the second term on the RHS we have switched $\omega \leftrightarrow \omega'$.  Furthermore, this second term is now exactly in the form of the original term, so we have the result:
\begin{eqnarray}
\sum_{\omega, \omega'} \left[A_\omega \tilde{\rho} A_{\omega'}^\dagger \mathcal{B}_{\omega \omega'} + A_{\omega'} \tilde{\rho} A_{\omega}^\dagger \left(\mathcal{B}_{\omega \omega'} \right)^{\ast} \right] &=&\frac{1}{2} \sum_{\omega, \omega'} \left[A_{\omega} \tilde{\rho} A_{\omega'}^\dagger b_{\omega \omega'} + A_{\omega'} \tilde{\rho} A_{\omega}^\dagger b_{\omega' \omega}  \right]  = \sum_{\omega, \omega'} A_{\omega} \tilde{\rho} A_{\omega'}^\dagger b_{\omega \omega'} \ .
\end{eqnarray}
For the remaining terms, we write:
\bes
\begin{align}
\sum_{\omega, \omega'} \left[A_{\omega'}^\dagger  A_\omega \tilde{\rho} \mathcal{B}_{\omega \omega'} + \tilde{\rho} A_{\omega}^\dagger A_{\omega'}   \left(\mathcal{B}_{\omega \omega'} \right)^{\ast} \right] & = \frac{1}{2} \sum_{\omega, \omega'} \left[ A_{\omega'}^\dagger  A_\omega \tilde{\rho} b_{\omega \omega'} + \tilde{\rho} A_{\omega}^\dagger A_{\omega'}  b_{\omega' \omega} \right] \nonumber \\ 
& + \frac{1}{2} \sum_{\omega, \omega'} \left[ A_{\omega'}^\dagger  A_\omega \tilde{\rho} \left( \mathcal{B}_{\omega \omega'} - \left(\mathcal{B}_{\omega' \omega} \right)^{\ast} \right) +  \tilde{\rho} A_{\omega}^\dagger  A_{\omega'} \left( \mathcal{B}_{\omega' \omega} - \left( \mathcal{B}_{\omega \omega'} \right)^{\ast} \right)  \right]  \\
& =  \frac{1}{2} \sum_{\omega, \omega'}  \left(b_{\omega \omega'} +  \mathcal{B}_{\omega \omega'} - \left( \mathcal{B}_{\omega' \omega} \right)^{\ast} \right) \left[ A_{\omega'}^\dagger  A_\omega \tilde{\rho}  + \tilde{\rho} A_{\omega'}^\dagger A_{\omega}  \right] \ .
\end{align}
\ees
Putting all this together we can now go from Eq.~\eqref{eqt:TimeIndependent1} to Eq.~\eqref{eq:K2-gen}.

\section{Relation between the Coarse-Grained Master Equation and the Standard Lindblad Equation} \label{appendix:RWA}

In order to study the relationship between the CG-SME and the RWA-SME, we begin with the dissipative coefficients of the CP map:
\begin{equation}
 b_{\omega \omega'}(t)=\intop_0^{t}ds\intop_0^{t}ds'e^{i(\omega's-\omega s')}\mathcal{B}(s,s') \ .
\end{equation}
We switch to new variables $u=s-s'$ and $v=s+s'$ using the homogeneity of the bath correlation function:
\begin{equation}
 b_{\omega \omega'}(t)=\frac{1}{2}\intop_0^{t}dve^{i\frac{\omega'-\omega}{2} v}\intop_{-v}^{v}du e^{i\frac{\omega+\omega'}{2}u}\mathcal{B}(u,0)+\frac{1}{2} \intop_{t}^{2 t}dve^{i\frac{\omega'-\omega}{2} v}\intop_{-(2t-v)}^{2 t-v}du e^{i\frac{\omega+\omega'}{2}u}\mathcal{B}(u,0) \ .
\end{equation}
We replace $v$ by $2t-v$ in the latter integral:
\begin{eqnarray}
 b_{\omega \omega'}(t)&=&\frac{1}{2}\intop_0^{t}dve^{i\frac{\omega'-\omega}{2} v}\intop_{-v}^{v}du e^{i\frac{\omega+\omega'}{2}u}\mathcal{B}(u,0)+ \frac{1}{2} e^{i\frac{\omega'-\omega}{2}2t}\intop_{0}^{t}dve^{-i\frac{\omega'-\omega}{2} v}\intop_{-v}^{v}du e^{i\frac{\omega+\omega'}{2}u}\mathcal{B}(u,0)\\
&=&e^{i\frac{\omega'-\omega}{2}t}\intop_0^{t}dv\cos\left(\frac{\omega'-\omega}{2}(v-t)\right)\intop_{-v}^{v}due^{i\frac{\omega+\omega'}{2}u}\mathcal{B}(u,0) \ .
\end{eqnarray}
For $\omega=\omega'$ this expression reads
\begin{equation}
 b_{\omega \omega}(t)=\intop_0^tdv\intop_{-v}^{v}due^{i\omega u}\mathcal{B}(u,0) \ ,
\end{equation}
and partial integration yields
\bes
\begin{align}
 b_{\omega\omega}(t)=&-\intop_0^{t}dv v\left( e^{i\omega v}\mathcal{B}(v,0)+e^{-i\omega v}\mathcal{B}(-v,0)\right)+t \intop_{-t}^{t}du e^{i\omega u}\mathcal{B}(u,0)\\ 
 =&\  t\intop_{-t}^{t}du\left(1-\frac{u}{2 t}\right)e^{i\omega u}\mathcal{B}(u,0) \ .
\end{align}
\ees
For our SME, we are interested in $b_{\omega \omega'}(\Delta t) / \Delta t$.  Setting $\Delta t \to \infty$, the part proportional to $\frac{u}{2t}\mathcal{B}(u,0)$ is negligible, and we recover the RWA result of Eq.~\eqref{eqt:RWA}:
\begin{equation}
\lim_{\Delta t \to \infty} \frac{1}{\Delta t}b_{\omega \omega}(\Delta t)=\intop_{-\infty}^{\infty}due^{i\omega u}\mathcal{B}(u,0)=\gamma(\omega) \ .
\end{equation}
For $\omega\not=\omega'$ we also perform a partial integration, but this time the boundary terms vanish:
\begin{equation}
 b_{\omega \omega'}(t)=-\frac{2 e^{i\frac{\omega'-\omega}{2}t}}{(\omega'-\omega)}\intop_0^{t}dv\sin\left( \frac{(\omega'-\omega)}{2}(v-t)\right)\left[e^{i\frac{\omega+\omega'}{2}v}\mathcal{B}(v,0)+e^{-i\frac{\omega+\omega'}{2}v}\mathcal{B}(-v,0)\right] \ .
\end{equation}
Changing from $v$ to $-v$ in the second term we get
\bes
\begin{align}
 b_{\omega \omega'}(t)=&-\frac{2 e^{i\frac{\omega'-\omega}{2}t}}{(\omega'-\omega)}\Bigg[\intop_0^{t}dv\sin\left(\frac{(\omega'-\omega)}{2}(v-t)\right)e^{i\frac{\omega+\omega'}{2}v}\mathcal{B}(v,0) \nonumber \\
+&\intop_{-t}^{0}dv\sin\left(\frac{(\omega'-\omega)}{2} (-v-t)\right)e^{i\frac{\omega+\omega'}{2}v}\mathcal{B}(v,0)\Bigg]\\
=&\frac{e^{i\frac{\omega'-\omega}{2}t}}{(\omega'-\omega)}\intop_{-t}^{t}dv\left[\sin\left(\frac{\omega'-\omega}{2}t\right)\left(e^{i \omega v}+e^{i \omega' v}\right)+\frac{\mathrm{sgn(v)}}{i}\cos\left(\frac{\omega'-\omega}{2}t\right)\left(e^{i \omega v}-e^{i \omega' v}\right)\right]\mathcal{B}(v,0) \ ,
\end{align}
\ees
where we used the angle sum identity for the sine in the last equality. If we again look at the term  $b_{\omega \omega'}(\Delta t) / \Delta t$ and take the limit of $\Delta t \to \infty$, we get
\beq
\lim_{\Delta t \to \infty} \frac{b_{\omega \omega'}(\Delta t)}{\Delta t} = \lim_{\Delta t \to \infty} \frac{e^{i\frac{\omega'-\omega}{2}\Delta t}}{(\omega'-\omega)\Delta t}\left[\sin\left(\frac{\omega'-\omega}{2}\Delta t\right)(\gamma(\omega)+\gamma(\omega'))+2\cos\left(\frac{\omega'-\omega}{2}\Delta t\right)(S(\omega)-S(\omega'))\right] \ .
\eeq
where we have used that for $\Gamma(\omega) = \int_0^{\infty} ds e^{i \omega s} \mathcal{B}(s,0)$, we have $\gamma(\omega) = \Gamma(\omega) + \Gamma(\omega)^{\ast}$ and $2 i S(\omega) = \Gamma(\omega) - \Gamma(\omega)^\ast$.  Since nothing cancels with the overall $\Delta t^{-1}$, we find that the $\omega \neq \omega'$ term vanishes, which is what we would expect for the RWA result.  A similar calculations could be done for the Lamb shift term \eqref{eq:Lambshift}.  Therefore, the RWA results can be understood as the $\Delta t \to \infty$ limit of the coarse-graining parameter.

\section{Calculation for the Two Level Example} \label{app:2level}

We consider the model in section \ref{sec:trivial}.  Let us denote the eigenvalues of $H_S$ by $\varepsilon_{\pm} = \pm \omega_0 / 2$ and their respective eigenvectors by:
\begin{equation}
| \varepsilon_+ \rangle \to \left( 1 , 0 \right)^{T}  \ , \quad | \varepsilon_- \rangle \to \left( 0  ,  1 \right)^{T}\ ,
\end{equation}
and $^T$ denotes the transpose. Note that the Lindblad operators in both the CG-SME and RWA-SME are given by:
\begin{eqnarray}
A_{-\omega_0} &=& | \varepsilon_+ \rangle \langle \varepsilon_+ | \sigma_z | \varepsilon_- \rangle \langle \varepsilon_- | = 0 \\
A_0 &=&  | \varepsilon_+ \rangle \langle \varepsilon_+ | \sigma_z | \varepsilon_+ \rangle \langle \varepsilon_+ |  +  | \varepsilon_- \rangle \langle \varepsilon_- | \sigma_z | \varepsilon_- \rangle \langle \varepsilon_- | = \sigma_z \\
A_{\omega_0} &=& | \varepsilon_- \rangle \langle \varepsilon_- | \sigma_z | \varepsilon_+ \rangle \langle \varepsilon_+ | = 0 
\end{eqnarray}
so in this two state model, only the $\omega= 0$ term contributes to the sums over $\omega$.  This in turn gives that $H_{\mathrm{LS}} \propto \ident$ such that $\left[ H_{\mathrm{LS}},\rho_S(t) \right] = 0$.  In order to solve the SMEs, it is convenient to vectorize the master equation.  Let us define:
\begin{equation}
\vec{\rho} \equiv \mathrm{vec}(\rho) = \left(
\begin{array}{c}
\rho_{i1} \\
\vdots \\
\rho_{i 2} \\
\vdots
\end{array}
\right)
\end{equation}
where $\mathrm{vec}(\rho)$ corresponds to stacking the columns of $\rho$ (in some basis). We now use the identity \cite{ZAMM:ZAMM19920721231}:
\begin{equation}
\mathrm{vec}\left( A B C \right) = \left( C^{T} \otimes A \right) \mathrm{vec}\left(B \right)
\end{equation}
where $(A,B,C)$ are arbitrary matrices, such that we can write both SME as
\begin{eqnarray}
\mathrm{vec}\left( \frac{d}{dt} \tilde{\rho}_S \right)&=& \gamma \left( A_{0}^{\ast} \otimes A_{0} -\frac{1}{2} \left(  \ident^T \otimes A_{0}^\dagger A_{0} + \left(A_{0}^\dagger A_{0}  \right)^{T} \otimes \ident  \right) \right) \vec{\rho}_S  \equiv \mathcal{L} \vec{\rho}_S(t)
\end{eqnarray}
%
where it is to be understood that $\gamma = \gamma_{\textrm{CG}}$ for the CG-SME and $\gamma = \gamma_{\textrm{RWA}}$ for the RWA-SME.  These are defined as:
\beq
\gamma_{\textrm{CG}} = \frac{1}{\Delta t}  \int_0^{\Delta t} ds \int_0^{s} ds' \left( B(s,s') + B(s',s) \right) \ ,\quad \gamma_{\textrm{RWA}} = \int_{-\infty}^{\infty} d s  B(s,0)
\eeq
The solution written in terms of the superoperator $\mathcal{L}$ is simply 
\begin{equation}
\vec{\rho}_S(t) = \exp( \mathcal{L} t) \vec{\rho}_S(0) 
\end{equation}
Using that $A_0 = \sigma_z$, $\mathcal{L}$ reduces to :
\begin{equation}
\mathcal{L}= \gamma \left( \sigma_z \otimes \sigma_z - \ident \otimes \ident \right)
\end{equation}
Conveniently, $\mathcal{L}$ is diagonal with entries $(0, -2 \gamma, -2 \gamma, 0)$, so we can immediately write:
\begin{equation}
\exp(\mathcal{L}) = \left( \begin{array}{cccc}
1 &&& \\
& \exp(-2 \gamma t) && \\
&& \exp(-2 \gamma t )& \\
&&& 1
\end{array} \right)
\end{equation}
Therefore, we find as a final result that:
\begin{equation}
\tilde{\rho}_S(t) = \left( \begin{array} {cc}
\tilde{\rho}_{11}(0) & \exp(-2 \gamma t) \tilde{\rho}_{12}(0) \\
\exp(-2 \gamma t) \tilde{\rho}_{21}(0) & \tilde{\rho}_{22}(0) 
\end{array} \right)
\end{equation}
Moving to the Schr\"odinger picture, the result is adjusted to
\begin{equation}
\rho_S(t) = \left( \begin{array} {cc}
\tilde{\rho}_{11}(0) & \exp(-2 \gamma t- i \omega_0 t) \tilde{\rho}_{12}(0) \\
\exp(-2 \gamma t + i \omega_0 t) \tilde{\rho}_{21}(0) & \tilde{\rho}_{22}(0) 
\end{array} \right)\ .
\end{equation}
Finally, for the infinite harmonic oscillator bath to bath operator $B = \lambda \sum_k (b_k + b_k^{\dagger} )$, we can explicitly calculate the form of the $\gamma$'s by using the result for the two point correlation:
\begin{equation}
\langle B(s) B(s') \rangle = \langle B(s - s') B(0) \rangle =\lambda^2 \int_0^{\infty} d \omega \frac{J(\omega)}{1 - e^{-\beta \omega} } \left( e^{-i \omega (s-s')} + e^{i \omega (s-s') - \beta \omega} \right)
\end{equation}
This then gives:
\bes
\begin{align}
\gamma_{\textrm{CG}} = & 2 \lambda^2 \frac{1}{\Delta t} \int_0^{\infty} J(\omega) \coth \left( \frac{\beta \omega}{2} \right) \left( \frac{ 1 - \cos(\omega \Delta t) }{\omega^2} \right)  = \lambda^2 \pi \int_0^{\infty} J(\omega)  \coth \left( \frac{\beta \omega}{2} \right)  \frac{\Delta t \ \mathrm{sinc}^2 \left( \frac{\omega \Delta t}{2} \right)}{\pi} \\
\gamma_{\textrm{RWA}} = & 2 \pi \lambda^2 \lim_{\omega \to 0} \left( \frac{ J(\omega)}{1-e^{- \beta \omega}} \right)
\end{align}
\ees
When $\gamma_{\textrm{CG}}$ is written in the form of the second equality, it becomes clear that taking the $\Delta t \to \infty$ limit corresponds to the RWA result since we can use the identity:
\beq
\lim_{\Delta t \to \infty} \Delta t \frac{\mathrm{sinc}^2 \left( \frac{\omega \Delta t}{2} \right)}{\pi} = \delta (\omega)
\eeq 
and $\gamma_{\textrm{CG}}$ becomes $\gamma_{\textrm{RWA}}$.
%
\section{Calculations for the Three Level Example for the Averaged Evolution Generator}\label{sec:AEcalc}
%
Here we provide the detailed derivations of the expressions for the averaged evolution generator of the three level model in section \ref{sec:model}. Let us calculate Eq.~\eqref{eq:bomompr} for the zero temperature case with the correlation function Eq.~\eqref{eq:zeroTcorr}:
\bea
b_{\omega \omega'}(t) &=& \int_0^t  d s \int_0^t  d s'  e^{i ( \omega' s - \omega s' )}\intop_0^\infty d \nu e^{-i\nu (s-s')}J(\nu)\nonumber \\
&=&\intop_0^\infty d \nu J(\nu)\int_0^t  d s  e^{i s(\om'-\nu)}\int_0^t  d s' e^{-i s' (\om-\nu)}\nonumber\\
&=&\intop_0^\infty d \nu \frac{J(\nu)}{(\om-\nu)(\om'-\nu)}\left(e^{i(\om'-\nu)t}-1\right)\left(e^{-i(\om-\nu)t}-1\right)\nonumber\\
&=&\intop_0^\infty d \nu \frac{J(\nu)}{(\om-\nu)(\om'-\nu)}\left(e^{i(\om'-\om)t}-e^{i(\om'-\nu)t}-e^{-i(\om-\nu)t}+1\right).
\eea
For the second equality sign we used the fact that the Fourier transform of the spectral function converges absolutely for a well behaved bath:
\beq
\intop_0^\infty d \nu \left|e^{-i\nu (s-s')}J(\nu)\right|=\intop_0^\infty d \nu \left|J(\nu)\right|<\infty \ .
\eeq
Let us look at the case $\om\not=\om'$. With the help of
\beq\label{eq:partfrac}
\frac{1}{(\om-\nu)(\om'-\nu)}=\frac{1}{\om'-\om}\left(\frac{1}{\om-\nu}-\frac{1}{\om'-\nu}\right) \ ,
\eeq
we can split up the integral:
\bea 
b_{\omega \omega'}(t) &=& \frac{1}{\om'-\om}\intop_0^\infty d \nu J(\nu)\Bigg[\frac{e^{i(\om'-\om)t}}{\om-\nu}-\frac{e^{i(\om'-\om)t}}{\om'-\nu}-\frac{e^{i(\om'-\nu)t}}{\om-\nu}+\frac{e^{i(\om'-\nu)t}}{\om'-\nu}-\frac{e^{-i(\om-\nu)t}}{\om-\nu}+\frac{e^{-i(\om-\nu)t}}{\om'-\nu}+\frac{1}{\om-\nu}-\frac{1}{\om'-\nu}\Bigg]\nonumber\\
&=& \frac{1}{\om'-\om}\left[\left(1+e^{i(\om'-\om)t}\right)\left(I_1(\om)-I_1(\om')\right)+e^{i\om' t}\left(I_2^*(\om')-I_2^*(\om)\right)+e^{-i\om t}\left(I_2(\om')-I_2(\om)\right)\right]\nonumber\\
&&
\eea
where we have defined the integrals
\beq 
\label{eq:ints}
I_1(\om)=\intop_0^\infty  d \nu \frac{J(\nu)}{\om-\nu}  \ , \quad I_2(\om)=\intop_0^\infty  d \nu \frac{J(\nu)e^{i\nu t}}{\om-\nu} \ ,
\eeq
For the ohmic spectral density \eqref{eq:ohmic} the integrals $I_1$ and $I_2$ are analytically solvable:
\bea
I_1(\om)&=&\eta\intop_0^\infty  d \nu \frac{\nu e^{-\frac{\nu}{\om_c}}}{\om-\nu}=-\eta\intop_{-\om}^\infty  d \nu \frac{(\nu+\om) e^{-\frac{\nu+\om}{\om_c}}}{\nu}=-\eta\om_c+\eta\om e^{-\frac{\om}{\om_c}}\text{Ei}\left(\frac{\om}{\om_c}\right)\nonumber \ , \\
&&\\
I_2(\om)&=&\eta\intop_0^\infty  d \nu \frac{\nu e^{-\nu\left(\frac{1}{\om_c}-i t\right)}}{\om-\nu}=-\eta\intop_{-\om}^\infty  d \nu \frac{(\nu+\om) e^{-(\nu+\om)\left(\frac{1}{\om_c}-i t\right)}}{\nu}\nonumber\\
&=&-\frac{\eta\om_c}{1-i\om_c t}+\eta\om e^{-\om\left(\frac{1}{\om_c}-i t\right)}\text{Ei}\left(\om\left(\frac{1}{\om_c}-i t\right)\right) \ ,
\eea
where we substituted $\nu$ by $\nu+\om$ in both calculations and used the definition \eqref{eq:Ei}.
For the case $\om=\om'$ we get from \eqref{eq:simpbwwpr}, again substituting $\nu$ by $\nu+\om$
\bea
b_{\om\om}(t)&=&2\intop_0^\infty  d \nu \frac{J(\nu)}{(\om-\nu)^2}(1-\cos((\om-\nu)t)) = 2\intop_{-\om}^\infty d \nu \frac{J(\nu +\om)}{\nu^2}(1-\cos(\nu t)) \ .
\eea
For the ohmic spectral function this yields
\bea
b_{\om\om}(t)&=&2\eta\intop_{-\om}^\infty d \nu \frac{(\nu +\om)e^{-\frac{\nu+\om}{\om_c}}}{\nu^2}(1-\cos(\nu t))\nonumber\\
&=&\eta e^{-\frac{\om}{\om_c}}\left[\text{Ei}\left(\frac{\om}{\om_c}+i\om t\right)+\text{Ei}\left(\frac{\om}{\om_c}-i\om t\right)-2 \text{Ei}\left(\frac{\om}{\om_c}\right)\right]+2\eta\om e^{-\frac{\om}{\om_c}} \intop_{-\om}^\infty d\nu\frac{e^{-\frac{\nu}{\om_c}}}{\nu^2}\left(1-\cos(\nu t)\right) \ .
\eea
Partial integration simplifies the last integral to
\bea
&&\intop_{-\om}^\infty d\nu\frac{e^{-\frac{\nu}{\om_c}}}{\nu^2}\left(1-\cos(\nu t)\right)=\intop_{-\om}^\infty d\nu\frac{e^{-\frac{\nu}{\om_c}}}{\nu}\left(t\sin(\nu t)-\frac{1}{\om_c}(1-\cos(\nu t))\right)+\frac{e^{\frac{\om}{\om_c}}}{\om}(1-\cos(\om t))\nonumber\\
&=&\frac{1}{\om_c}Ei\left(\frac{\om}{\om_c}\right)-\frac{1}{2}\left(\frac{1}{\om_c}+i t\right)\Ei\left(\frac{\om}{\om_c}+i\om t\right)-\frac{1}{2}\left(\frac{1}{\om_c}-i t\right)\Ei\left(\frac{\om}{\om_c}-i\om t\right)+\frac{e^{\frac{\om}{\om_c}}}{\om}(1-\cos(\om t)) \ ,
\eea
so that we can write altogether
\bea
b_{\om\om}(t)&=&\eta e^{-\frac{\om}{\om_c}}\left[\left(1-\frac{\om}{\om_c}-i \om t\right)\Ei\left(\frac{\om}{\om_c}+i \om t\right)+\left(1-\frac{\om}{\om_c}+i \om t\right)\Ei\left(\frac{\om}{\om_c}-i \om t\right)+2\left(\frac{\om}{\om_c}-1\right)\Ei\left(\frac{\om}{\om_c}\right)\right] \nonumber \\
&& +2\eta(1-\cos(\om t)) \ .
\eea

Let us now calculate the Lamb shift term for the zero temperature bath correlation function Eq.~\eqref{eq:zeroTcorr}. Similar to the case of $b_{\om \om'}$ we get
\bea 
S_{\om \om'}(t)&=&-\frac{i }{2}\intop_0^\infty  d \nu J(\nu)\intop_0^t d s\intop_0^s  d s' \left( e^{i (\omega' s - \omega s') } e^{-i \nu (s-s')} - e^{-i (\omega s - \omega' s') } e^{-i \nu (s'-s)}\right)\nonumber\\
&=&-\frac{i }{2}\intop_0^\infty  d \nu J(\nu)\intop_0^t d s\left[e^{i s(\om'-\nu)}\intop_0^s  d s'e^{i s'(\nu-\om)}- e^{i s(\nu-\om)}\intop_0^s  d s'e^{i s'(\om'-\nu)}\right]\nonumber\\
&=&-\frac{i }{2}\intop_0^\infty  d \nu J(\nu)\intop_0^t d s\left[e^{i s(\om'-\nu)}\frac{e^{i s(\nu-\om)}-1}{i(\nu-\om)}+e^{i s(\nu-\om)}\frac{e^{i s(\om'-\nu)}-1}{i(\nu-\om')}\right]\nonumber\\
&=&-\frac{ i}{2}\intop_0^\infty  d \nu J(\nu)\intop_0^t d s\left[\frac{e^{i s(\om'-\om)}-e^{i s(\om'-\nu)}}{i(\nu-\om)}+\frac{e^{i s(\om'-\om)}-e^{i s(\nu-\om)}}{i(\nu-\om')}\right].
\eea
Now we have to distinguish the two cases, $\om =\om'$ and $\om\not=\om'$. For $\om\not=\om'$ we get
\bea
S_{\om \om'}(t)&=&-\frac{ i}{2}\intop_0^\infty  d \nu J(\nu)\Bigg[\frac{e^{i t(\om'-\om)}-1}{(\om-\om')(\nu-\om)}-\frac{e^{-i t(\nu-\om')}-1}{(\nu-\om)(\nu-\om')}+\frac{e^{i t(\om'-\om)}-1}{(\om-\om')(\nu-\om')}+\frac{e^{i t(\nu-\om)}-1}{(\nu-\om)(\nu-\om')}\Bigg].\nonumber\\
&=&-\frac{ i}{2}\intop_0^\infty  d \nu J(\nu)\Bigg[\frac{e^{i t(\om'-\om)}-1}{(\om-\om')(\nu-\om)}+\frac{e^{i t(\om'-\om)}-1}{(\om-\om')(\nu-\om')}+\frac{e^{i t(\nu-\om)}-e^{-i t(\nu-\om')}}{(\nu-\om)(\nu-\om')}\Bigg] \ .
\eea
With the help of \eqref{eq:partfrac} and \eqref{eq:ints} this formula simplifies to
\beq
S_{\om \om'}(t)=-\frac{i}{2(\om-\om')}\left[\left(1-e^{i t(\om'-\om)}\right)\left(I_1(\om)+I_1(\om')\right)+e^{-i\om t}\left(I_2(\om')-I_2(\om)\right)+e^{i\om' t}\left(I_2^*(\om)-I_2^*(\om')\right)\right] \ .
\eeq
For the case $\om=\om'$ Eq.~\eqref{eq:Spartres} reads
\bea
S_{\om \om}(t)&=&-\frac{ i}{2}\intop_0^\infty  d \nu J(\nu)\intop_0^t d s\left[\frac{1-e^{i s(\om-\nu)}}{i(\nu-\om)}+\frac{1-e^{i s(\nu-\om)}}{i(\nu-\om)}\right]\nonumber\\
&=&-\frac{ i}{2}\intop_0^\infty  d \nu J(\nu)\left[\frac{2 t}{i(\om-\nu)}+\frac{e^{i(\nu-\om)t}-e^{-i(\nu-\om)t}}{(\om-\nu)^2}\right]\nonumber\\
&=&\intop_0^\infty  d \nu J(\nu)\left[\frac{ t}{\nu-\om}+\frac{\sin((\nu-\om)t)}{(\nu-\om)^2}\right]\nonumber\\
&=&- t I_1(\om)+I_3(\om) \ ,
\eea
where $I_1$ is defined in \eqref{eq:ints} and $I_3$ is defined as
\beq
I_3(\om)=\intop_0^\infty  d \nu J(\nu)\frac{\sin((\nu-\om)t)}{(\nu-\om)^2} \ .
\eeq
For the Ohmic spectral density this yields
\bea
I_3(\om)&=&\eta\intop_0^\infty  d \nu \nu e^{-\frac{\nu}{\om_c}}\frac{\sin((\nu-\om)t)}{(\nu-\om)^2} = \eta e^{-\frac{\om}{\om_c}}\intop_{-\om}^\infty  d u (u+\om) e^{-\frac{u}{\om_c}}\frac{\sin(ut)}{u^2}\nonumber\\
&=&\eta\left[\frac{e^{-\frac{\om}{\om_c}}}{2 i}\left(\Ei\left(\frac{\om}{\om_c}+i\om t\right)-\Ei\left(\frac{\om}{\om_c}-i\om t\right)\right)+\om e^{-\frac{\om}{\om_c}}I_4(\om)\right] \ ,
\eea
where we defined
\beq
I_4(\om)=\intop_{-\om}^\infty  d u\frac{e^{-\frac{u}{\om_c}}\sin(ut)}{u^2} \ .
\eeq
Partial integration yields
\bea
I_4(\om)&=&-\intop_{-\om}^\infty  d u\frac{e^{-\frac{u}{\om_c}}\left(\frac{1}{\om_c}\sin(ut)+t\cos(ut)\right)}{u}+\frac{e^{\frac{\om}{\om_c}}\sin(\om t)}{\om}\nonumber\\
&=&-\intop_{-\om}^\infty  d u\frac{e^{-\frac{u}{\om_c}}\left[e^{itu}\left(\frac{1}{i\om_c}+t\right)+e^{-itu}\left(-\frac{1}{i\om_c}+t\right)\right]}{2u}+\frac{e^{\frac{\om}{\om_c}}\sin(\om t)}{\om}\nonumber\\
&=&\left(t+\frac{1}{i\om_c}\right)\Ei\left(\frac{\om}{\om_c}-i\om t\right)+\left(t-\frac{1}{i\om_c}\right)\Ei\left(\frac{\om}{\om_c}+i\om t\right)+\frac{e^{\frac{\om}{\om_c}}\sin(\om t)}{\om} \ .
\eea
Using this result, we get the final result for $I_3$:
\bea
I_3(\om)&=& \eta\frac{e^{-\frac{\om}{\om_c}}}{2 i}\left[\left(1-2\frac{\om}{\om_c}+2i\om t\right)\Ei\left(\frac{\om}{\om_c}+i\om t\right)-\left(1-2\frac{\om}{\om_c}-2 i\om t\right)\Ei\left(\frac{\om}{\om_c}-i\om t\right)\right]+\eta\sin(\om t) \ .
\eea

\end{document}